\documentclass[a4paper,12pt]{article}
\usepackage[latin1]{inputenc}
\usepackage{amsmath,amssymb,array,graphicx,cite}

\makeatletter
\normalsize
\newcommand\ZZ{\mathbb{Z}}

\newcommand\QQ{\mathbb{Q}}
\newcommand{\CC}{\mathbb{C}}
\newcommand{\PP}{\mathbb{P}}
\newcommand\beq{\begin{equation}}
\newcommand\eeq{\end{equation}}
\newcommand\beqn{\begin{eqnarray}}
\newcommand\eeqn{\end{eqnarray}}
\def\eq#1{(\ref{#1})}
\def \keti{\rangle\!\rangle}

\def \brai{\langle\!\langle}

\def \ie{{\it i.e.}~}
\def \eg{{\it e.g.}~}
\def \cf{{\it cf.}~}
\def \su{\widehat{\mathfrak{su}}}
\def \u{\widehat{\mathfrak u}}

\def\p{{}^\prime}
\def\wt#1{\widetilde{#1}}
\def\id{{\rm id}}
\newcommand{\Ext}{\mathrm{Ext}}
\newcommand{\Hom}{\mathrm{Hom}}
\newcommand{\cM}{{\cal M}}
\newcommand{\HH}{{\cal H}}
\newcommand{\VV}{{\cal V}}
\newcommand{\II}{{\cal I}}
\newcommand{\N}[3]{{\cal N}_{{#1} {#2}}^{\phantom{#1}\;{#3}}}
\newcommand{\ol}{\overline}
\newcommand{\0}{\Omega}
\newcommand{\tr}{{\rm tr}}

\newcommand{\lra}{\to}

\newcommand{\we}{\wedge}

\def\lrdelp{\overrightarrow{\partial_+}}
\def\lrdelm{\overrightarrow{\partial_-}}
\def\ts{\textstyle}
\def\math{\mathsurround=0pt}
\def\overrightarrow#1{\vbox{\math\ialign{##\crcr
  $\leftrightarrow$\crcr\noalign{\kern-1pt\nointerlineskip}
   \hfil$\displaystyle{#1}$\hfil\crcr}}}

\def\oh{{\ts {1\over2}}}

\DeclareMathOperator{\im}{im}
\DeclareMathOperator{\coker}{coker}
\DeclareMathOperator{\ch}{ch}
\DeclareMathOperator{\rk}{rk}

\renewcommand\section{\@startsection {section}{1}{\z@}%
                                   {-3.5ex \@plus -1ex \@minus -.2ex}%
                                   {2.3ex \@plus.2ex}%
                                   {\normalfont\large\bfseries}}
\renewcommand\subsection{\@startsection{subsection}{2}{\z@}%
                                     {-3.25ex\@plus -1ex \@minus -.2ex}%
                                     {1.5ex \@plus .2ex}%
                                     {\normalfont\normalsize\bfseries}}
\makeatother

\begin{document}
  
\begin{titlepage}

\title{Permutation branes \\ and linear matrix factorisations}

\vskip1.3cm
\author{H\aa kon~Enger$^a$, Andreas~Recknagel$^b$,\\
        Daniel~Roggenkamp$^c$\\
\\[3mm]
\normalsize\sl ${}^a$Department of Physics, University of Oslo,\\
\normalsize\sl P.\ O.\ Box 1048 Blindern, NO-0316 Oslo, Norway\\
\small \tt hakon.enger@fys.uio.no\\[.7\baselineskip]
\normalsize\sl ${}^b$King's College London, Department of Mathematics,\\
\normalsize\sl Strand, London WC2R 2LS, UK\\
\small \tt anderl@mth.kcl.ac.uk\\[.7\baselineskip]
\normalsize\sl ${}^c$Department of Mathematical Sciences, University of Durham, \\
\normalsize\sl South Rd, Durham DH1 3LE, UK\\
{\small \tt daniel.roggenkamp@durham.ac.uk}
}

\date{ }

\maketitle

\vspace{.8cm}

\begin{abstract}
  All the known rational boundary states for Gepner models can be 
  regarded as permutation branes. On general grounds, one expects 
  that topological branes in Gepner models can be encoded as matrix
  factorisations of the corresponding Landau-Ginzburg potentials. 
  In this paper we identify the matrix factorisations associated to arbitrary
  B-type permutation branes. 
\end{abstract}

\vfill

\thispagestyle{empty}
\end{titlepage}

\section{Introduction and outline}

The study of strings and D-branes on Calabi-Yau spaces is a remarkably 
rich area. These string compactifications are interesting for 
phenomenological reasons (in the heterotic string version they come 
closest to realistic particle spectra; if D-branes are added to the
type II version, they lead to $N=1$ low-energy theories), as well as 
for mathematical reasons: In the 1980s, string theorists conjectured 
the existence of mirror symmetry for Calabi-Yau target spaces, which has 
since been refined, by including D-branes, to what is often called the 
homological mirror symmetry programme, involving derived categories 
of coherent sheaves and the Fukaya category on the target manifolds
\cite{Konts:1994}. 

These conjectures deal with the large-volume regime of string theories on 
CY target spaces, but they were at least partially inspired by investigations of 
the stringy regime, where efficient descriptions in terms $N=2$ world-sheet 
theories are available, most notably in the form of Gepner models (orbifolds 
of tensor products of $N=2$ superminimal models). 
The connection to sigma-models on Calabi-Yau manifolds can be established 
via Landau-Ginzburg models \cite{wi93}, the critical points of which 
have a description in terms of minimal models. 

Even though Gepner models \cite{gep88} are defined in a somewhat abstract way, 
they are rational conformal field theories, so it is possible to find 
symmetry-preserving boundary conditions (\ie D-branes) for them
\cite{Recknagel:1998sb}. Using those boundary states and relating 
open string Witten indices to geometrical intersection forms, the authors 
of  \cite{Brunner:1999jq} achieved a large-volume interpretation of the 
CFT boundary conditions; see also \cite{DiaRoem,KLLW,Scheid:2000} for further 
studies along these lines. 
Since minimal models are closely related to $N=2$ Landau-Ginzburg models, 
the study of supersymmetric boundary conditions for the latter should 
from the outset be relevant to the study of D-branes on Calabi-Yau manifolds. 
Special (linear) boundary conditions for LG models were analysed 
in \cite{gov1,gov2} and also in \cite{hiv}. 
Later, building on early work by Warner \cite{Warner:1995ay}, a connection 
between LG boundary conditions and factorisations of the respective LG 
superpotential into two matrices was established in \cite{bhls03,Kapustin:2002bi,kl03a}. This was 
motivated by an unpublished proposal due to M.\  Kontsevich, who suggested 
that topological B-branes in LG models can be described by matrix
factorisations of the respective LG potential, as discussed in great detail in the 
papers by Orlov \cite{Orlov:2003yp} and Kapustin and Li 
\cite{Kapustin:2002bi,kl03a,kl03b}.  

From any $N=2$ superconformal field theory one can obtain two
$2$-dimensional topological quantum field theories, the A- and the
B-model, by performing the respective twists and restricting the full 
Hilbert space to the cohomology of the BRST operator, which provides 
the `physical states' of the topological model. This is also true
for non-conformal $N=2$ models with unbroken $R$-symmetries, \eg 
for LG models with affine target space and quasi-homogeneous
superpotentials (see \eg \cite{hvms}), which will be studied in this article.

In B-models of such LG theories on world-sheets with boundary,
the matrix factors of the bulk superpotential determine 
the boundary conditions, and in particular the boundary BRST operators \cite{bhls03},  
therefore the spectrum of physical open string states. 

Significantly, the main pieces of CFT data used by Brunner, Douglas et al.\ 
to extract large-volume information in \cite{Brunner:1999jq} were all taken 
from the `topological sector' of the Gepner model: In particular, the 
intersection form counts open string Ramond ground states (with fermion 
number), so is precisely given by the Euler number of the cohomology of 
the boundary BRST operator.

\vspace{.3cm}

There are some important questions in string theory which are, at present, 
too hard to answer directly in the CFT, but can be tackled in the simplified 
framework of topological theories. In particular, one can study deformation 
away from the ``Gepner point'', induced by marginal bulk (and boundary) 
fields. The properties of topological D-branes in deformed backgrounds can 
be encoded in topological D-brane superpotentials; a mainly 
perturbative approach was presented in \cite{dougov}, recent progress 
towards computing exact superpotentials has been made in \cite{hll04a,hll04b}. These
superpotentials provide a very efficient description of D-brane stability, 
and of characteristic CFT data like the chiral ring structure. 

Thus it is of some interest to study which LG boundary conditions 
(given in the form of matrix factorisations) correspond to the 
CFT boundary conditions that are known for Gepner models so far.  
All these are rational boundary states,  and can be viewed as permutation 
branes, as introduced in \cite{r02}: Whenever $n$ of the minimal model constituents of the Gepner model have the same level, there is a 
non-trivial action of the permutation group $S_n$, which can be 
used to build boundary states that obey permutation gluing conditions, 
\ie where the left-moving super-Virasoro generators of the $i^{\rm th}$ 
minimal model are glued to the right-moving generators of the 
$\sigma(i)^{\rm th}$ model for any permutation $\sigma\in S_n$.

Landau-Ginzburg matrix factorisations that reproduce the topological 
spectra of D-branes in $N=2$ minimal models are well-known 
\cite{Kapustin:2002bi,kl03a,kl03b,h04}, 
and from these one can form (orbifolds of) tensor products which 
reproduce the spectra of the simplest $\sigma =\;$id `permutation' branes 
from \cite{Recknagel:1998sb}, see \eg  \cite{Ashok:2004zb} for a detailed 
discussion. The next-simplest permutation branes involving length two 
cycles were analysed in \cite{Brunner:2005fv}, where the connection between 
CFT boundary states and the `rank one matrix factorisations' discussed in 
\cite{Ashok:2004zb} was established by computations of spectra and comparison 
of various other CFT and LG results. In the present paper, we propose 
a correspondence between arbitrary permutation branes and a special 
class of {\sl linear matrix factorisations}, which 
were studied from a purely algebraic point of view in 
\cite{Backelin:1988}. A linear matrix factorisation is a decomposition 
of a homogeneous (degree $d$) polynomial $W(x_1,\ldots,x_n) =
\alpha_0\cdots\alpha_{d-1}$ into $d$ matrices, each of which 
is linear in the $x_i$. Grouping some $\alpha_i$ together, one obtains 
two-term matrix factorisations $W = p_0 p_1$. The general construction we
propose covers the cases discussed before (trivial permutation or cycles of at most 
length two), but it necessarily involves higher rank matrix factors as 
soon as the permutation has cycles of length three or more. 

The evidence we present in support of the proposed correspondence is
partly in the form of computer-algebraic computations for explicit 
matrix factorisations, leading to topological spectra (in particular
to Witten indices) which are compared 
to results obtained in the corresponding Gepner models; here we make 
extensive use of the package Macaulay2 \cite{M2}. In addition, 
we present a general derivation of the BRST cohomology 
for open strings stretching between an
arbitrary permutation brane and special $\sigma =\;$id branes (tensor 
products of minimal model boundary conditions). Employing 
tools from homological algebra a bit more ingeniously, it should be 
possible to extend this calculation to arbitrary tensor product branes, 
but already the special cases considered here should be a sufficient 
starting point to compute charges for arbitrary permutation branes. 

The body of the paper starts with a review of the relation between 
matrix factorisations and topological LG models; in particular, we 
spell out how the boundary BRST-cohomology is encoded in Ext-groups. 
In Section 3, we revisit boundary states in minimal models and Gepner 
models and derive the topological open string spectra of the permutation 
branes, which in particular yields the Witten index. The main new results are 
contained in Chapter 4: We first present linear matrix factorisations of 
Landau-Ginzburg potentials $W= x_1^d + \ldots + x_n^d$ in 
Section 4.1, then formulate a conjecture which of these correspond 
to the topological permutation branes from the third chapter.
Sections 4.2 and 4.3 contain evidence for
this correspondence. Some homological algebra arguments and the 
Macaulay2 codes together with results on the large-volume Chern characters for 
permutation branes in the $(k=3)^5$-Gepner model describing a
sigma-model on the quintic threefold in $\PP^4$ are collected in the Appendix.

\vspace{.3cm}
Apart from open technical problems like finding simpler and more general 
proofs for the correspondence between certain linear matrix factorisations and
rational Gepner model branes, there are some conceptual, and also 
some physical questions that would be interesting to study in the future. 
For example, one could try to exploit the concrete Ext-groups arising 
in our examples as a starting point of a computation of brane superpotentials. 
It should also be interesting to compare our description of permutation 
LG branes to the one given in \cite{EGJ:2005}, and to try and extend 
the present constructions to $D$-type modular invariants, see 
\cite{BG:2005b} for some recent results. 

On the whole, it is probably fair to say that the link between topological 
B-branes in LG models and matrix factorisations on the one hand and 
boundary CFT on 
the other is, as yet, rather loose, and that a deeper understanding of 
the connection would be desirable. For instance, only part of the
linear factorisations which are described in Section 4.1 actually
correspond to permutation branes, and one wonders what CFT
boundary conditions the additional factorisations correspond to -- if any. 
They might correspond to non-rational (symmetry-breaking) Gepner 
boundary states, which so far are not at all under control, and one may 
hope that matrix factorisations point towards new constructions of CFT 
boundary conditions. For these and other reasons, it is definitely 
worth-while to aim at a better understanding of the TFT-CFT interplay.

\section{Landau-Ginzburg models and matrix factorisations}

In this chapter, we briefly recall the relation between topological B-type 
branes in  Landau-Ginzburg models and  matrix factorisations,  and 
how tools from homological algebra can be used to describe 
data of topological string theory. 

An $N=2$ supersymmetric Landau-Ginzburg model with target space $\CC^n$
on a world-sheet $\Sigma$ is given by the {\sl bulk action} 
\begin{align}
S_{\Sigma} &=  \int_{\Sigma}\! d^2x\;\left[ 
  \partial^\mu\bar X^j\partial_\mu X^j - i\,\bar\psi^{j}_{-}\lrdelp\psi^{j}_{-}
  - i\,\bar\psi^j_{+}\lrdelm\psi^j_{+} \right.\nonumber\\
&\quad\qquad\qquad\ \ + \left.{\ts {1\over4}}\,|\partial W|^2 + \oh\, W_{ij}\psi^i_{+}\psi^j_{-}
   + \oh\, \overline{W}_{ij} \bar\psi^i_{-}\bar\psi^j_{+}\right]
\end{align}
where $X^j$, $1\leq j\leq n$, are bosonic fields, $\psi^j_{\pm}$ left- and right-moving 
fermions, $W(X)$ is the Landau-Ginzburg potential, 
and $W_{ij} :=  \partial^2 W/\partial X^i\partial X^j$. 
(The world-sheet carries the 2-d version of the `mostly minus' metric.)

This action is invariant under the diagonal $N=2$ supersymmetry
transformation as long as the world-sheet has no boundary; for
$\partial\Sigma\neq\emptyset$, one adds {\sl boundary terms} \cite{Warner:1995ay,bhls03}
\beqn
S_{\partial\Sigma, \psi} = {i\over4}\ \sum_{j}\ \int_{\partial\Sigma}dx^0\ 
\Bigl[\; \bar\theta^j \eta^j - \bar\eta^j\theta^j\;\Bigr]
\eeqn
(with $\eta := \psi_- + \psi_+$, $\theta := \psi_- - \psi_+$) as well
as a term involving additional boundary fermions $\pi_\alpha$, and boundary potentials
$p_i^\alpha(X)$
\begin{align}
&S_{\partial\Sigma, \pi} = \sum_{\alpha, j}\ \int_{\partial\Sigma}dx^0\ 
\Bigl[\; i\, \bar\pi^\alpha \partial_0\pi^\alpha - \oh\,\bar p_{0}^{\alpha}  p_{0}^{\alpha}
- \oh\,\bar p_{1}^{\alpha}  p_{1}^{\alpha}\phantom{mmmmmm}\nonumber\\
&\quad \qquad \ +\oh\,\pi^\alpha (\bar\eta^j\, \ol{\partial}_j\bar p_{0}^{\alpha} +i\eta^j\, \partial_j p_{1}^{\alpha}) 
-\oh\,\bar\pi^\alpha (\eta^j\,  \partial_jp_{0}^{\alpha} -i\bar\eta^j\, \ol{\partial}_j\bar p_{1}^{\alpha}) 
 \;\Bigr]
\end{align}
In order to preserve diagonal B-type $N=2$ supersymmetry, the
potentials $p_{i}^{\alpha}$ (taken to be polynomial in the $X^j$) have to satisfy
the factorisation condition  \cite{bhls03}
$$
\sum_\alpha  p_{0}^{\alpha}\,p_{1}^{\alpha}  = W 
$$
(up to a possible additive constant on the rhs, which will be set to
zero in the following).

These potentials also determine the action of the (boundary
contribution to the) {\sl BRST operator},
$$
Q\; X = 0 \ ,\quad Q\; \pi = p_0\ ,\quad Q\; \bar\pi = -i p_1\ \ . 
$$
In the topological field theory, physical open string states
correspond to cohomology classes of $Q$.

The  space ${\cal P}$ on which the boundary fields act is graded by the fermion number, 
${\cal P} = {\cal P}_0 \oplus {\cal P}_1$, and using 
Clifford algebra anticommutation relations among the boundary fermions 
$$
\{\,\pi_\alpha , \,\pi_\beta\} = \{\,\bar\pi_\alpha , \,\bar\pi_\beta\} = 0\ ,\quad 
\{\,\pi_\alpha, \,\bar\pi_\beta\} = \delta_{\alpha,\beta}
$$
one can view $Q$ as acting  on boundary fields 
$$
\Phi\, = \begin{pmatrix}  f_{00} & f_{10} \cr f_{01} & f_{11}\cr \end{pmatrix},
$$
where $f_{ij} : {\cal P}_i \to {\cal P}_j$, by graded commutator with the matrix
$$
\Theta\, = \begin{pmatrix}  0 &p_1 \cr p_0 & 0\cr \end{pmatrix}
$$
(actually, for $\alpha = 1,\ldots,r$, $Q$ is a $2^r \times 2^r$ matrix).

It is straightforward to carry this over to strings stretching between two 
different branes (where $\Phi: {\cal P} \lra \wt{\cal P}$ and
$Q\,\Phi = 
\Theta\, \Phi \pm \Phi\, \wt \Theta$). Furthermore, one can generalise this 
view of the BRST cohomology by allowing for matrices $p_0, p_1$ of 
arbitrary size, see \cite{laz}. In this way, while losing an explicit realisation through 
a Clifford algebra spanned by LG boundary fermions $\pi_\alpha$, one 
makes contact to general {\sl matrix factorisations}, which are pairs of  square 
matrices $p_i \in {\rm Mat}(k,A)$ over the polynomial ring $A=\CC[X_j]$ such that 
$$
p_0\;p_1 =  W\; {\bf 1}_k\ \ .
$$
Note that the physical content of a matrix factorisation is invariant under 
gauge transformations as formulated \eg in \cite{Hori:2004ja,w04}:
Two matrix factorisations $(p_0,p_1)$ and $(p_0',p_1')$ are called 
{\sl equivalent} if there are two invertible matrices 
$U,V \in {\rm GL}(k,A)$ 
with 
\begin{equation}\label{equivMF}
U\,p_0\,V^{-1} = p_0'\quad{\rm and}\quad V\,p_1\,U^{-1} = p_1'\ \ .
\end{equation} 
Therefore boundary conditions in topological LG models are indeed
described by equivalence classes of matrix factorisations. This will
be understood implicitly in the following.

\vspace{.3cm}

A simple way to obtain boundary conditions in certain LG models is by 
means of the {\sl tensor product construction}. Whenever the superpotential 
$W$ is a sum of two polynomials in different variables
$W(x_1,\ldots,x_n)=W_1(x_1,\ldots,x_m)+W_2(x_{m+1},\ldots,x_n)$ then the 
LG model with superpotential $W$ is indeed 
a tensor product of the two LG models with superpotentials $W_1$, $W_2$.
Therefore it must be possible in this situation to choose boundary conditions
in each of these models separately to obtain the ``product'' boundary 
condition in the LG model with potential $W$. It turns out 
that the corresponding matrix factorisation 
is the tensor product matrix factorisation: Let $(p_0,p_1)$, $(q_0,q_1)$ 
be matrix factorisations of $W_1$ and $W_2$, respectively, 
then the tensor product of these is given by the pair of matrices
\begin{equation}\label{tpfactorisation}
  r_0 =
\begin{pmatrix}
  p_0 \otimes 1 & -1 \otimes q_1 \\
  1 \otimes q_0 & p_1 \otimes 1
\end{pmatrix},
\quad
  r_1 =
\begin{pmatrix}
  p_1 \otimes 1 & 1 \otimes q_1 \\
  -1 \otimes q_0 & p_0 \otimes 1
\end{pmatrix}.
\end{equation}
As discussed in \cite{Ashok:2004zb}, $(r_0,r_1)$ indeed
gives rise to the open string spaces associated to tensor product
boundary conditions.

\vspace{0.3cm}

Invoking some basic notions from homological algebra, 
we can relate the spaces
of topological open string states, \ie the cohomology of the
BRST-operator Q, to certain {\sl Ext-groups}, which will
prove useful for calculations later on. To this end, to a matrix
factorisation $(p_0,p_1)$ of $W$ of rank $k$, we associate the
$A$-module $P=\coker(p_1)$ and its $A$-free resolution
\beq\label{Aresolution}
0 \longrightarrow A^k\stackrel{p_1}{\longrightarrow}A^k\longrightarrow
P\longrightarrow 0\,.
\eeq
Given another matrix factorisation $(\wt p_0,\wt p_1)$ of $W$ of rank
$\wt k$, we obtain another module $\wt P=\coker \wt p_1$ in the same way.
It is easy to see that the space of 
bosonic BRST-cocycles associated to the pair of matrix factorisations
$(p_0,p_1)$ and $(\wt p_0,\wt p_1)$ is isomorphic to the space of
chain maps between the respective resolutions \eq{Aresolution}. 
(A chain map between two complexes $(C_n,\partial_n)$ and $(\wt
C_n,\wt\partial_n)$ is given by a 
sequences of maps $f_n: C_n \to \wt C_{n}$ satisfying 
$f_{n-1}\;\partial_n = \wt \partial_{n}\;f_n$.) If the two
complexes are resolutions of $C$ and $\wt C$ respectively, one can show 
that the space of homomorphisms $\Hom(C,\wt C)$ is isomorphic to the 
space of chain maps between the respective resolutions modulo the space 
of chain homotopies. (A homotopy between two chain maps $f$ and 
$f\p$ is a sequence of maps $h_n : C_n \lra \wt C_{n+1}$ satisfying 
$f_n - f\p_n = h_{n-1} \;\partial_{n} + \wt\partial_{n+1}\; h_{n}$.) 
However,  one can check that the space of chain homotopies between 
resolutions \eq{Aresolution} is only a subspace of the image of the 
BRST-operator $Q$ (see also below), essentially because these 
resolutions are ``too short". Thus the bosonic part of the 
BRST-cohomology in general is a quotient of $\Hom_A(P,\wt P)$. 

To obtain a better description of the BRST-cohomology, one can use the
fact that due to $W\,P=0$, $P$ is also a module over the ring
$R=A/(W)$. Since $p_1p_0=W\,\id_{A^k}=p_0p_1$, this module has
the $2$-periodic $R$-free resolution
\begin{equation}\label{Rresolution}
  \cdots \longrightarrow R^{k} \stackrel{p_1}{\longrightarrow} R^{k}  
  \stackrel{p_0}{\longrightarrow} R^{k} 
  \stackrel{p_1}{\longrightarrow} R^{k}  
  \longrightarrow P \longrightarrow 0\,.
 \end{equation}
This is a complex with $\partial_{2n} = p_0$ 
and $\partial_{2n-1} = p_1$ for all $n\geq 1$. 

Resolutions (whether periodic or not) can be used to calculate the 
groups $\Ext^i_R(P,\cdot)$. Namely, for two 
modules $M$ and $N$ over a ring $S$, $\mathrm{Ext}^i_S(M,N)$
is defined to be the $i^{\rm th}$ right derived functor of the functor $\Hom_S(\cdot,N)$,
\ie given a projective resolution 
$\cdots \stackrel{\partial_3}{\to}  
M_2 \stackrel{\partial_2}{\to} M_1 \stackrel{\partial_1}{\to} M_0 \to M \to 0$ of $M$, it can be calculated
as the $i^{\rm th}$ cohomology of the complex
\begin{equation}\label{homcomplex}
  0 \longrightarrow \mathrm{Hom}_S(M_0,N) \longrightarrow \mathrm{Hom}_S(M_1,N) \longrightarrow \cdots,
\end{equation}
where the maps are induced by the maps $\partial_i$ in the
resolution of $M$, namely 
$f_i \in \mathrm{Hom}_S(M_i,N) \mapsto 
f_i \circ \partial_{i+1} \in \mathrm{Hom}_S(M_{i+1},N)$.

A perhaps more concrete way to represent $\Ext$-groups is (see \eg \cite{Hilton:1971}):
\begin{eqnarray}
&\mathrm{Ext}_S^0(M,N) = \mathrm{Hom}_S(M,N)\ \ , \\
&\mathrm{Ext}_S^i(M,N)  = {\rm coker}\bigl(
  \mathrm{Hom}_S(M_{i-1},N) \longrightarrow   \mathrm{Hom}_S(K_i,N)\bigr)
\end{eqnarray}
where $K_i := \im\partial_{i} \subset M_{i-1}$. 

We can, however,  use \eq{homcomplex} directly to make contact with 
the cohomology of the BRST-operator $Q$ associated to matrix
factorisations $(p_0,p_1)$ and $(\wt p_0,\wt p_1)$ of W: In 
even degree, $\ker Q$ is isomorphic
to the space of maps $f_{00}\in\Hom_R(R^k,R^{\wt k})$ such that
there exists an $f_{11}\in\Hom_R(R^k,R^{\wt k})$ with $f_{00} p_1=\wt
p_1f_{11}$. Likewise, the even degree part of $\im Q$ is isomorphic to
$\Hom_R(R^k,R^{\wt k})\circ p_0 + \wt p_1\circ\Hom_R(R^k,R^{\wt k})$.
Dividing out the second summand from $\ker Q$ means that we can choose
representatives for $f_{00}$, which are zero on $\im\wt p_1$. This is
achieved by passing from $\Hom_R(R^k,R^{\wt k})$ to $\Hom_R(R^k,\wt P)$
everywhere, and the condition to belong to $\ker(Q)$ becomes
$f_{00}\,p_1 =0$. In these representatives, the remaining part of
$\im(Q)$ is just given by $\Hom_R(R^k,\wt P)\circ p_0$ and 
one easily sees that $\ker(Q)/\im(Q)$ can be obtained 
as the even cohomology of the complex \eq{homcomplex} with $M=P$ and
$N=\wt P$. Thus,
the bosonic part of the BRST-cohomology is
isomorphic to $\Ext^{2i}_R(P,\wt P)$ for $i>0$. (Because of the 
two-periodicity of the resolution \eq{Rresolution}, all these Ext-groups 
are isomorphic.)

To obtain the odd part of the BRST-cohomology, one can replace $(p_0,p_1)$
by the shifted matrix factorisation $(-p_1,-p_0)$ in the discussion
above,  and one finds that the odd BRST-cohomology is isomorphic to  
$\Ext^{2i-1}_R(P,\wt P)$ for $i>0$. 

Altogether, we arrive at the statement that the spaces of states of 
bosonic respectively fermionic open strings in LG models with boundary conditions
characterised by matrix factorisations $(p_0,p_1)$, $(\wt p_0,\wt p_1)$ of $W$ are
given by
\beqn
\mathrm{H}^{\rm even}(Q) =
\mathrm{Ext}^{2i}_R(P,\wt P)\,,\;
\mathrm{H}^{\rm odd}(Q) =
\mathrm{Ext}^{2i-1}_R(P,\wt P)\nonumber\ \ 
\eeqn
for $i>0$, 
where $P=\coker p_1$ and $\wt P=\coker \wt p_1$ are the $R$-modules
obtained from the respective matrix factorisations.
Interchanging $p_0$ and $p_1$ amounts 
to switching to the anti-brane of $P$ and thus exchanging the notions
of bosons and fermions in the open string sectors.

This identification of BRST-cohomology with 
$\Ext$-groups of the modules $P$, $\wt P$ 
allows us to exploit the machinery of homological algebra (in particular long 
exact sequences in homology induced by short exact sequences of
modules) in the analysis of topological open strings in LG models.

Let us remark at this point that the modules $P$, $\wt P$ are rather special. 
Eisenbud \cite{Eisenbud:1980} (see also
\cite{Avramov:2001} for a slight generalisation) showed that {\sl all}
minimal free resolutions of finitely generated modules over polynomial
rings $R=\CC[x_1,\ldots,x_n]/(W)$ become 2-periodic after at most $n$
steps. 
The modules for which a
minimal free resolution is 2-periodic from the start -- 
exactly the ones
induced by matrix factorisations -- are the
{\sl maximal Cohen-Macaulay} modules.  These have been studied rather 
extensively in the mathematical literature.

\vspace{0.3cm}

One aspect of LG models which we have not mentioned up to now is
that they carry an {\sl action of a discrete group} $\Gamma$. Indeed, if the
superpotential $W$ is homogeneous of degree
$d$, which we will assume throughout the paper, this group
is given by $\Gamma=\ZZ_d$ and it acts on the bosonic fields by multiplication
with a primitive $d^{\rm th}$ root of unity $\xi$: 
$X_i\mapsto\xi^t X_i$ for $t\in\ZZ_d$. This also induces 
actions on the open string spaces, and the analysis of the
respective representations will be useful for the identification of
matrix factorisations associated to conformal boundary conditions. 

In terms of matrix factorisations, this group action can be 
formulated as follows \cite{Ashok:2004zb}: The $\ZZ_d$-action on 
the $X_i$ gives the ring $R$ the structure of a $\ZZ_d$-graded ring (\ie
the ring structure is compatible with the $\ZZ_d$-action), 
and one can consider $\ZZ_d$-graded modules over it.
The latter are modules $P$ over $R$ together with representations
$\rho:\ZZ_d\rightarrow\mathrm{End}(P)$ of $\ZZ_d$ on them, 
which are compatible with the module structure.
In particular, the maps $p_0$ and $p_1$ of a matrix factorisation
$\ \, p_1:P_1{\rightleftarrows} P_0:p_0\ \,$ can be taken
as maps between $\ZZ_d$-graded modules $(P_0,\rho_0)$ and $(P_1,\rho_1)$,
which also have to be compatible with the grading, \ie
$\rho_1(g)\circ p_0 = p_0 \circ \rho_0(g)$ and
$\rho_0(g)\circ p_1 = p_1 \circ \rho_1(g)$
for all $g\in\ZZ_d$. Pictorially we write
this as
\begin{equation}\label{equiv}
  {}\raisebox{2ex}{$\overset{\rho_1}{\rotatebox[origin=c]{-120}{$\circlearrowleft$}}$} P_1
  \overset{p_1}{\underset{p_0}{\rightleftarrows}}
  P_0 \raisebox{2ex}{$\overset{\rho_0}{\rotatebox[origin=c]{120}{$\circlearrowleft$}}$} \,.
\end{equation}
Such graded matrix factorisations then give rise to $\ZZ_d$-graded Ext-groups,
whose gradings specify the corresponding actions on the corresponding
open strings states.

Incorporation of the $\ZZ_d$-action not only provides finer
information about the boundary conditions in LG models, but 
also allows to carry the treatment of boundary
conditions in LG models over to boundary conditions in 
LG-orbifolds with orbifold group $\ZZ_d$. The effect 
of the orbifolding on the LG model is that the respective open string sectors 
are projected onto $\ZZ_d$-invariant subspaces. In terms of matrix
factorisations this means that 
the space of open 
strings in the LG orbifold model is given by the mod-$d$-degree-$0$ parts of the 
Ext-groups describing the corresponding spaces of open strings in the 
underlying LG model.

LG orbifolds are relevant because of their relation to non-linear sigma 
models: The $\ZZ_d$-orbifold of a LG model with homogeneous
superpotential $W$ of degree $d$ in $n$ variables corresponds 
to a non-linear sigma model defined on the hypersurface $X=\{W=0\}\subset\PP^{n-1}$ 
in projective space \cite{wi93}, as long as $X$ is a Calabi-Yau manifold, 
which in the situation considered here is the case if $n=d$. The
$\ZZ_d$-action in the LG model appears here as a ``remainder'' of the
$\CC^*$-action divided out to obtain the projective hypersurface.
In this
case one expects that B-type boundary conditions in the LG orbifold
also have a geometric interpretation as B-type D-branes in the non-linear sigma model
on $X$. The latter are believed to be described by objects in the 
bounded derived category of coherent sheaves ${\cal D}^b(\mathrm{Coh}(X))$ 
on $X$.

\section{LG boundary conditions at the conformal point}
\label{cftsection}

The critical behaviour of Landau-Ginzburg models with superpotential
$W=x_1^{k_1+2}+\ldots+x_n^{k_n+2}$ can be described in terms of tensor 
products 
\beq\label{tpmodel}
\cM_{k_1,\ldots,k_n}=\cM_{k_1}\otimes\ldots\otimes\cM_{k_n}
\eeq
of $N=2$ minimal models $\cM_k$ with $A$-type modular invariants. 

We are interested in B-type boundary conditions for those tensor product 
theories that preserve the $N=2$ supersymmetries of each of the minimal 
model, \ie satisfy gluing conditions for all of the $N=2$ super-Virasoro
algebras separately. Certainly, there are the obvious ones, namely 
``tensor products'' of boundary conditions of each of the $\cM_{k_i}$. 
However, if some of the factor models are isomorphic, \ie $k_i=k_j$ for 
some $i\neq j$, then it is also possible to construct boundary conditions 
whose gluing conditions permute the $N=2$ super-Virasoro algebras of the 
respective models. Such boundary conditions are called {\sl permutation 
boundary conditions}. 

For tensor products of rational CFTs with diagonal modular invariant there is a
standard construction for the corresponding permutation boundary states \cite{r02}. 
This construction has to be slightly modified when dealing with B-type gluing 
automorphisms, with respect to which  minimal models are not diagonal. 
(A somewhat pedestrian approach to tackle similar problems in constructing
permutation branes for Gepner models, \ie orbifolded tensor products of
minimal models, was employed in \cite{r02}.)

The minimal models $\cM_k$ are conformal field theories which are
rational with respect to the action of an $N=2$-super Virasoro algebra at central charge $c_k={3k\over k+2}$.
The bosonic part of this super Virasoro algebra can be realised as the coset W-algebra
$(\su(2)_k\oplus\u(1)_4)/\u(1)_{2k+4}$. In fact, the respective coset
model can be obtained from the $\cM_k$ by a non-chiral GSO-projection, see \eg \cite{bh03}. 

The Hilbert space of the $\cM_k$ can be decomposed into irreducible
highest weight representations of the respective super Virasoro
algebra. It is convenient however, to consider the decomposition into
irreducible highest weight representations $\VV_{[l,m,s]}$ of its
bosonic subalgebra
\beq\label{minmodhilbertspace}
\HH_k\cong\bigoplus_{[l,m,s]\in\II_k} \VV_{[l,m,s]}\otimes\left(\ol\VV_{[l,m,s]}\oplus
\ol\VV_{[l,m,s+2]}\right)\,,
\eeq
where the set of such representations is
\beq\label{repset}
\II_k=\{(l,m,s)\,|\,0\leq l\leq k\,,\,m\in\ZZ_{2(k+2)}\,,\,s\in\ZZ_4\,,\,l+m+s\in 2\ZZ\}/\sim
\eeq
with the field identification $(l,m,s)\sim(k-l,m+k+2,s+2)$.

Apart from the alignment of R- and NS-sectors, the Hilbert space of the tensor product
model \eq{tpmodel} is just given by the tensor product of the individual minimal model Hilbert 
spaces
\beq\label{tphilbertspace}
\HH_{k_1,\ldots,k_n}=\bigoplus_{\stackrel{[l_i,m_i,s_i]\in\II_{k_i}}
{s_i-s_j\in 2\ZZ}}\bigotimes_{i=1}^n \VV_{[l_i,m_i,s_i]}\otimes
\left(\ol\VV_{[l_i,m_i,s_i]}\oplus \ol\VV_{[l_i,m_i,s_i+2]}\right)\,.
\eeq
The model possesses the symmetry group $\ZZ_{k+2}\times\ZZ_2$ whose
generators $g\in\ZZ_{k+2}$ and $h\in\ZZ_2$ act on 
$\VV_{[l,m,s]}\otimes\ol\VV_{[l,\ol m,\ol s]}$
by multiplication with $e^{{\pi i\over k+2}(m+\ol m)}$ and
$e^{{\pi i\over 2}(s+\ol s)}$, respectively.

We would like to analyse boundary conditions whose gluing automorphisms
permute the $N=2$ algebras of the $\cM_{k_i}$ in \eq{tpmodel}, and since 
only isomorphic $N=2$ algebras can be ``glued together'', we will restrict 
ourselves to the case of tensor products of $n$ identical minimal models 
$\cM_k$, \ie $k_i=k$ for all $i$. 

We first review the construction for the trivial permutation $\sigma =\,$id, 
which is also discussed in great detail in \cite{bhw04}. Already in this case, 
where the gluing conditions factorise into $n$ independent ones, the 
corresponding boundary states are not just tensor products of single minimal 
model boundary states because of the sector alignment. Furthermore, 
one has to take into account that a single minimal model is not
diagonal with respect to the B-type gluing automorphism.

\subsection{Trivial permutation}

The B-type gluing automorphism $\tau_B$ of the $N=2$ superconformal algebra
induces an isomorphism $\VV_{[l,m,s]}\stackrel{\sim}{\rightarrow}\VV_{[l,-m,-s]}$
of the minimal model representations. Therefore a sector
\beq\label{sector}
\VV_{[l,m,s]}\otimes\ol\VV_{[l,m,\ol s]}\subset\HH_{k}
\eeq
in a single minimal model
gives rise to an Ishibashi state satisfying B-type gluing conditions iff
$[l,m,s]=\tau_B[l,m,\ol s]=[l,-m,-\ol s]$.
Thus, in a single minimal model there are Ishibashi states
\beq\label{minmodishibashi}
|[l,0,s]\keti_B\,,\quad{\rm for}\;{\rm all}\; 
[l,0,s]\in\II_k\,.
\eeq
Note however, that one can also introduce Ishibashi states
$|[l,m,s]\keti_B$ for $[l,m,s]\in\II_k$ with $m\neq 0\;{\rm mod}\;(k+2)$,
if one allows for twists with respect to the $\ZZ_{k+2}$-symmetry of the 
model, \cf \cite{bh03}. These additional Ishibashi states appear in the 
decomposition of twisted boundary states, whose existence can be understood 
as follows: Since the boundary states built from the Ishibashi states 
\eq{minmodishibashi} are invariant under the group $\ZZ_{k+2}$, the
latter also acts on the respective open string sectors. Thus we can insert 
a $\ZZ_{k+2}$-generator in a trace over an open sector, and by means of 
modular transformation this can be rewritten as an overlap of twisted 
boundary states\footnote{Even though these are not states in the Hilbert 
space of the original model, they can nevertheless be used to describe 
the corresponding correlation functions.}; for more details about this 
point see \eg \cite{bh02}. With $g$ being a generator of $\ZZ_{k+2}$, 
the $g^{-t}$-twisted boundary states are given by 
\beqn
\|[L,M,S]\keti_{B,t}&=&{1\over\sqrt{k+2}}\sum_{a\in\ZZ_{k+2}}\sum_{[l,m,s]\in\II_k}
e^{{2\pi i\over k+2}a(m-t)}
{S_{[L,M,S],[l,m,s]}\over\sqrt{S_{\0,[l,m,s]}}}|[l,m,s]\keti_B\nonumber\\
&=&{1\over\sqrt{k+2}}\sum_{a\in\ZZ_{k+2}}\sum_{[l,m,s]\in\II_k}
e^{-{2\pi i\over k+2}at}
{S_{[L,M+2a,S],[l,m,s]}\over\sqrt{S_{\0,[l,m,s]}}}|[l,m,s]\keti_B\nonumber\\
&=&\sqrt{k+2}
\sum_{[l,t,s]\in\II_k}
{S_{[L,M,S],[l,t,s]}\over\sqrt{S_{\0,[l,t,s]}}}|[l,t,s]\keti_B\,,\label{minmodbd}
\eeqn
where in the last line it is summed over all $l,s$ such that
$[l,t,s]\in\II_k$, and 
\beqn\label{smatrix}
S_{[L,M,S],[l,m,s]}&=&{e^{-{i\pi\over 2}Ss}\over\sqrt{2}}{e^{{i\pi\over k+2}Mm}\over\sqrt{k+2}}
S_{L,l}\;\;\;\;{\rm and}\\
S_{L,l}&=&\sqrt{2\over k+2}\sin\left({\pi \over k+2}(L+1)(l+1)\right)\nonumber
\eeqn
are the modular $S$-matrices of the $N=2$ minimal models and the
$\su(2)_k$-WZW models respectively, and where $\0=[0,0,0]$ denotes the 
minimal model vacuum representation. In an untwisted boundary state, 
all the twisted Ishibashi states are projected out, and the label $M$
determines the representations of $\ZZ_{k+2}$ on the open string Hilbert
spaces.

The spectra of open string states with corresponding boundary conditions 
can easily be obtained from the overlaps of the respective boundary states. 
The bosonic part of open string states with boundary conditions corresponding 
to $\|[L,M,S]\keti_B$ and $\|[L\p,M\p,S\p]\keti_B$ is described by the 
overlap of $\|[L,M,S]\keti_B$ and $\|[L\p,M\p,S\p]\keti_B$, whereas the 
fermionic part of the spectrum is determined by the overlap of
$\|[L,M,S]\keti_B$ with the boundary state $\|[L\p,M\p,S\p+2]\keti_B$,
which is obtained from $\|[L\p,M\p,S\p]\keti_B$ by reversing the sign of 
its RR-part. 
Insertion of a power of the generator $g$ of the symmetry group $\ZZ_{k+2}$
in the trace over the open sector is achieved by considering the overlap
of the corresponding twisted boundary states. The calculation of the spectra
is straightforward and one obtains
\beqn
&&\tr_{\HH_{[L\p,M\p,S\p],[L,M,S]}^{\rm bos}}\!\!\left(g^{t} q^{L_0-{c\over 24}}\right)
\!=\!
{}_{-t,B}\brai[L\p,M\p,S\p]\|q^{{1\over 2}(L_0+\ol L_0)-{c\over 24}}\|[L,M,S]\keti_{B,-t}
\nonumber\\
&&\qquad\qquad=\sum_{\stackrel{[l,m,s]\in\II_k}{a\in\ZZ_{k+2}}}
\N{[L\p,M\p,S\p]}{[L,M+2a,S]}{[l,m,s]}\;
e^{{2\pi i\over k+2}ta}\;\chi_{[l,m,s]}(q)\,\label{minmodspectrum}
\eeqn
where $\chi_{[l,m,s]}$ are the characters and 
${\cal N}$ the fusion rules of the minimal model. 

Because of the sector alignment, one cannot obtain boundary conditions in
tensor products of minimal models just by tensoring the boundary conditions 
\eq{minmodbd} of single minimal models. Instead one has to project the tensor 
products of minimal model boundary states onto the contributions coming from 
Ishibashi states that are twisted with respect to the alignment group 
$\ZZ_2^{n-1}$. The result can be written as
\beqn\label{tpbd}
&&
\|L_1,\ldots,L_n,S_1,\ldots,S_n,M={\textstyle \sum_i} M_i\keti_{B,t}^\id\\
&&\qquad=2^{1-n\over 2}\!\!\!\!\!\!\sum_{b_2,\ldots,b_{n}\in\ZZ_2}\!\!\!
\|[L_1,M_1,S_1+2{\textstyle \sum_i} b_i]\keti_{B,t}\otimes
\bigotimes_{i=2}^{n}\|[L_i,M_i,S_i-2b_i]\keti_{B,t}
\nonumber\\
&&\qquad={(2k+4)^{n\over 2}\over \sqrt{2}}
\sum_{\stackrel{[l_i,t,s_i]\in\II_k}{s_i-s_j\in 2\ZZ}}
\prod_{i=1}^n{S_{[L_i,M_i,S_i][l_i,t,s_i]}\over\sqrt{S_{\0[l_i,t,s_i]}}}\bigotimes_{i=1}^n
|[l_i,t,s_i]\keti_B\,,\nonumber
\eeqn
where now $t$ refers to the twist by $g^{-t}$ with $g$ the generator of the 
diagonal $\ZZ_{k+2}\subset\ZZ_{k+2}^n$ of the product
of minimal model symmetry subgroups; as above, in the last line 
the sum is understood to be taken over $l_i,\, s_i$ such that $[l_i,t,s_i]\in\II_k$. 
Note that the boundary state \eq{tpbd} only depends on $M=\sum_iM_i$, which 
again determines the $\ZZ_{k+2}$-representations in the open sectors.
Namely, taking into account the form of the modular $S$-matrix \eq{smatrix},
we see that the boundary states depend on the $M$-labels only through a phase
$e^{{i\pi t\over k+2}(\sum_i M_i)}$ multiplying the $t$-twisted Ishibashi states.
Using \eq{minmodspectrum}, the spectrum of open strings with boundary
conditions corresponding to 
$\|\alpha\keti=\|{\scriptstyle L_1,\ldots,L_n,S_1,\ldots,S_n,M={\textstyle \sum_i}M_i}\keti_{B}^\id$
and 
$\|\alpha\p\keti=\|{\scriptstyle L\p_1,\ldots,L\p_n,S\p_1,\ldots,S\p_n,M\p={\textstyle \sum_i}M\p_i}\keti_{B}^\id$
can be easily determined
\beqn\label{tpspectrum}
&&\tr_{\HH_{\alpha\p \alpha}^{\rm bos}}\left(g^{t}q^{L_0-{c\over 24}}\right)={}_{-t}\brai\alpha\p\|
q^{{1\over 2}(L_0+\ol L_0)-{c\over 24}}\|\alpha\keti_{-t}\\
&&\qquad=\sum_{\stackrel{a_i\in\ZZ_{k+2}}{b_i\in\ZZ_2}}\sum_{[l_i,m_i,s_i]\in\II_k}
e^{{2\pi i\over k+2}t\sum_i a_i} \prod_{i=1}^n\chi_{[l_i,m_i,s_i]}(q)\nonumber\\
&&\qquad\qquad\times
\N{[L\p_1,M\p_1,S\p_1]}{[L_1,M_1+2a_1,S_1+2\sum_i b_i]}{[l_1,m_1,s_1]}
\prod_{i=2}^n \N{[L\p_i,M\p_i,S\p_i]}{[L_i,M_i+2a_i,S_i-2b_i]}{[l_i,m_i,s_i]}\,.\nonumber
\eeqn
The respective fermionic spectrum can be obtained from the bosonic one
by shifting an odd number of $S\p_i$ by 2.
Formula \eq{tpspectrum}  gives the expected spectrum for tensor 
product boundary conditions.  Namely, the sector alignment, \ie the sum over the $b_i$, 
ensures that the corresponding space of bosonic (fermionic) open strings is given by the 
tensor product of all combinations of bosonic and an even (odd) number of fermionic 
open string spaces of the individual models.

After this short review of product boundary conditions in tensor products of
minimal models let us return to permutation boundary conditions.

\subsection{Non-trivial permutation}

In this section we will present B-type boundary conditions in $\cM_k^{\otimes n}$ which preserve
all the individual $N=2$ algebras of the minimal models in a different
manner. Namely we impose gluing conditions which relate the holomorphic
algebra of the $i^{\rm th}$ minimal model to the antiholomorphic one of the 
$\sigma(i)^{\rm th}$ minimal model, where $\sigma\in S_n$ is a permutation. 

Since every permutation can be written as a product of cyclic permutations, 
we restrict our discussion to cyclic permutations $\sigma:(1,\ldots,n)
\mapsto(2,\ldots,n,1)$ of the $n$ factor models.
The treatment can easily be carried over to the general situation.

To construct boundary states satisfying $\sigma$-permuted B-type gluing conditions,
we first of all determine the respective Ishibashi states. 
The $t$-twisted sector
\beq
\VV_{[l_1,m_1,s_1]}\otimes\ldots\otimes\VV_{[l_n,m_n,s_n]}\otimes 
\ol\VV_{[l_1,m_1-2t,\ol s_1]}\otimes\ldots\otimes\ol\VV_{[l_n,m_n-2t,\ol s_n]}\subset\HH_{k,\ldots,k}^t
\eeq
gives rise to an Ishibashi state with respect to the $\sigma$-permuted B-type gluing automorphism iff
\beqn
[l_i,m_i,s_i]=\tau_B[l_{i+1},m_{i+1}-2t,\ol
s_{i+1}]=[l_{i+1},-m_{i+1}+2t,-\ol s_{i+1}]
\eeqn
for all $i\in\ZZ_n$. We can choose representatives so that this
condition becomes
\beq
l_i=l_1\,,\; 
m_{2i+1}=m\,,\; 
m_{2i}=2t-m\,,\;\;
\ol s_i=-s_{i+1}\;\; {\rm for\ all}\;\; i\,.
\eeq
For $n$ {\sl odd}, we obtain $m_i=t$ for all $i$, whereas for $n$
even, there are more solutions, namely $m_{2i+1}=m$ for all $i$ and
$m_{2i}=2t-m$. Thus, for {\sl odd} $n$ there are $t$-twisted Ishibashi
states 
\beq\label{noddIshibashi}
|[l,t,s_1,\ldots,s_n]\keti^\sigma\quad {\rm for\ all}\;\; [l,t,s_1,\ldots,s_n]\in \II_{k,n}\,,
\eeq
where
$\II_{k,n}=\{(l,m,s_1,\ldots,s_n)\,|\, [l,m,s_i]\in\II_k\}/\sim$ with
$(l,m,s_1,\ldots,s_n)\sim(k-l,m+k+2,s_1+2,\ldots,s_n+2)$.
For {\sl even} $n$ on the other hand there exist $t$-twisted Ishibashi
states 
\beq\label{nevenIshibashi}
|[l,m,2t-m,s_1,\ldots,s_n]\keti^\sigma\quad  {\rm for\ all}\;\; [l,m,s_1,\ldots,s_n]\in \II_{k,n}\,.
\eeq
Because of this difference between the case of B-type permutation 
boundary conditions involving permutations of even and odd cycle length, 
we will treat them separately in the following.

\subsubsection{Odd cycle length}

Our ansatz for the $\sigma$-permuted B-type boundary states is an
adaption of the permutation boundary states for diagonal CFTs \cite{r02} 
to our situation. We have to account for the absence of untwisted
Ishibashi states with $m\neq 0$, which can be done similarly to the 
case of a single minimal model discussed above. Furthermore we have to take 
care of the fact that the minimal models are non-diagonal. The alignment 
condition is automatically satisfied for $\sigma$-permuted gluing conditions. 
We define $t$-twisted boundary states as follows 
\beqn\label{noddbd}
&&\|L,M,S_1,\ldots,S_n\keti_{B,t}^\sigma\\
&&\qquad:={1\over\sqrt{k+2}}\sum_{a\in\ZZ_{k+2}}
\sum_{\stackrel{[l,m,s_1]\in\II_k}{s_i-s_1\in 2\ZZ}}e^{-{2\pi i\over k+2}at}\;
{S_{[L,M+2a,S_1][l,m,s_1]} \over \left(S_{\0[l,m,s_1]}\right)^{n\over
    2}} \;
{e^{-{i\pi\over 2}\sum_{i>1}S_is_i}\over 2^{n-1\over 2}}\nonumber\\
&&\qquad\qquad\qquad\qquad\qquad\qquad\qquad\qquad\qquad\quad
\times \;|[l,m,s_1,\ldots,s_n]\keti_B^\sigma\nonumber\\[2mm]
&&\qquad=
\sqrt{k+2}
\sum_{\stackrel{[l,t,s_1]\in\II_k}{s_i-s_1\in 2\ZZ}}
{S_{[L,M,S_1][l,t,s_1]} \over \left(S_{\0[l,t,s_1]}\right)^{n\over
    2}}\;
{e^{-{i\pi\over 2}\sum_{i>1}S_is_i}\over 2^{n-1\over 2}}\;
|[l,t,s_1,\ldots,s_n]\keti_B^\sigma\nonumber
\eeqn
Using standard facts about modular $S$-matrices, it is easy to obtain
the spectrum of open strings between two such permutation boundary 
states on both sides. The computation closely parallels the situation 
of permutation boundary conditions in diagonal CFTs \cite{r02}, one 
merely has to take into account the additional Ishibashi states with $s_i\neq s_j$
and the corresponding phases $e^{-{i\pi\over 2}\sum_{i>1}S_is_i}$ 
in the boundary states. To deal with them, we parametrise 
the $[l,m,s_1,\ldots,s_n]\in\II_{k,n}$ as $[l,m,s_1,s_1+b_2,\ldots,s_n+b_n]$,
where $[l,m,s_1]$ runs over $\II_k$ and $b_i\in\ZZ_2$ are arbitrary.
The sum over $b_2,\ldots,b_n\in\ZZ_2$ is independent of $s_1$, and 
the result for the open string spectrum between two $\sigma$-permutation 
branes $\|\alpha\keti=\|{\scriptstyle L,M,S_1,\ldots,S_n}\keti_{B}^\sigma$
and 
$\|\alpha\p\keti=\|{\scriptstyle L\p,M\p,S\p_1,\ldots,S\p_n}\keti_{B}^\sigma$ 
is
\beqn\label{noddspectrum}
&&\tr_{\HH_{\alpha\p \alpha}^{\rm bos}}\left(g^{t}q^{L_0-{c\over 24}}\right)={}_{-t}\brai\alpha\p\|
q^{{1\over 2}(L_0+\ol L_0)-{c\over 24}}\|\alpha\keti_{-t}\\
&&\qquad\qquad=\sum_{a\in\ZZ_{k+2}}
\sum_{[l_i,m_i,s_i]\in\II_k}
e^{{2\pi i\over k+2}at} 
\prod_{i=1}^n\chi_{[l_i,m_i,s_i]}(q)\nonumber\\
&&\qquad\qquad\qquad\qquad\qquad\times
\prod_{i=2}^n\delta^{(2)}_{s_i,S\p_i-S_i}\N{[L\p,M\p,{\scriptstyle \sum_i} S\p_i]}{[L,M+2a,{\scriptstyle
    \sum_i}S_i]}{[l_1,m_1,s_1]*\ldots*[l_n,m_n,s_n]}\,,
\nonumber
\eeqn
where $*$ denotes fusion of minimal model representations; the fusion rules ${\cal N}$
are extended linearly to sums of representations. Shifting
an odd number of $S$-labels of one of the boundary states by $2$ 
yields the fermionic spectrum.

Apart from the open sectors with $\sigma$-permuted B-type boundary conditions on
both sides, we are also interested in the ones with $\sigma$-permuted boundary
conditions on one and non-permuted boundary conditions on the other side.
For this, we first of all need the overlaps between the corresponding
$t$-twisted Ishibashi states:
\beqn\label{sigmaidoverlap}
&&{}_B\brai[l_1\p,t\p,s_1\p]|\otimes\ldots\otimes{}_B\brai[l_n\p,t\p,s_n\p]|
q^{{1\over 2}(L_0+\ol L_0)-{c\over 24}}|[l,t,s_1,\ldots,s_n]\keti^\sigma_B\qquad\\
&&\qquad\qquad\qquad\qquad\qquad\qquad
= \prod_{i=1}^n\left(\delta_{l_i\p,l}\,\delta_{s_i\p,s}\,\delta_{s_i,s}\right)
\tr_{\VV_{l,t,s}^{\otimes n}}\left(\sigma q^{L_0-{c\over 24}}\right)\,,\nonumber
\eeqn
where $\sigma$ acts on the tensor product space by permuting the factors.
As was noted in \cite{Brunner:2005fv} for the case $n=2$, 
this trace equals the character $\chi_{[l,t,s]}(q^n)$ only
up to a phase due to the (not necessarily bosonic) statistics
of the respective states. More precisely
\beq\label{twcharacter}
\chi_{[l,t,s]}(q^n)=\tr_{\VV_{l,t,s}^{\otimes n}}\left((-1)^{(1-n)F}\sigma q^{L_0-{c\over 24}}\right)
=e^{(n-1)\left({\pi it\over k+2}-{\pi i s\over 2}\right)}\tr_{\VV_{l,t,s}^{\otimes n}}\left(\sigma q^{L_0-{c\over 24}}\right)\,.
\eeq
Here, $(-1)^F=e^{2\pi iJ_0}$, where $J_0$ denotes the zero mode of the 
$U(1)$-current from the $N=2$ algebra. Using this and the modular 
transformation properties of the minimal model characters
\beq
\chi_{[l,t,s]}((Sq)^n)=\sum_{[l\p,t\p,s\p]\in\II_k}S_{[l,t,s][l\p,t\p,s\p]}\;
\chi_{[l\p,t\p,s\p]}\left(q^{1\over n}\right)\ \; ,
\eeq
the open string spectrum between a  $\sigma$-permuted boundary state 
$\|\alpha\keti=\|{\scriptstyle L,M,S_1,\ldots,S_n}\keti_{B}^\sigma$
on one side and a non-permuted tensor product boundary state $\|\alpha\p\keti=
\|{\scriptstyle L\p_1,\ldots,L\p_n,S\p_1,\ldots,S\p_n,M\p={\sum_i} 
M\p_i}\keti_{B}^\id$  on the other, follows as 
\beqn\label{sigmaidnodd}
&&\tr_{\HH_{\alpha\p \alpha}^{\rm bos}}\left(g^{t}q^{L_0-{c\over 24}}\right)={}_{-t}\brai\alpha\p\|
q^{{1\over 2}(L_0+\ol L_0)-{c\over 24}}\|\alpha\keti_{-t}\\
&&\qquad\qquad=(k+2)^{n-1\over 2}\sum_{[l,m,s]\in\II_k}\sum_{a\in\ZZ_{k+2}}
e^{{2\pi i\over k+2}at}\chi_{[l,m,s]}\left(q^{1\over n}\right)\nonumber\\
&&\qquad\qquad\qquad\qquad\times
\N{[L\p_1,M\p_1,S\p_1]*\ldots*[L\p_n,M\p_1,S\p_1]}{[L,M+2a+(1-n),{\sum_i}S_i+(1-n)]}{[l,m,s]}
\,.\nonumber
\eeqn
Note that the effect of the relative phase between twisted characters and $\sigma$-twisted
traces in \eq{twcharacter} on the open spectra \eq{sigmaidnodd} 
is a shift of coset-W-algebra representations $[l,m,s]\mapsto [l,m+(1-n),s+(1-n)]$ in the open channel.

\subsubsection{Even cycle length}

For even cycle length $n$, we define the $t$-twisted boundary states
\beqn\label{nevenbd}
&&\|L,M,T,S_1,\ldots,S_n\keti_{B,t}^\sigma:=
\sum_{[l,m,s_1,\ldots,s_n]\in\II_{k,n}}e^{{2\pi i\over k+2}Tt}\;
{S_{[L,M-2T,S_1],[l,m,s_1]}\over 
  \left(S_{\Omega,[l,m,s_1]}\right)^{n\over 2}}\qquad\\
&&\qquad\qquad\qquad\qquad\times
{e^{-{i\pi\over 2}\sum_{i=2}^n S_is_i}\over \sqrt{2}^{n-1}}\;
|[l,m,2t-m,s_1,\ldots,s_n]\keti^\sigma_B\ ; \nonumber
\eeqn
see also \cite{Brunner:2005fv} for the special case $n=2$. 
It is now straightforward to calculate the open string spectrum
between two such permutation boundary states.
For $\|\alpha\keti=\|{\scriptstyle L,M,T,S_1,\ldots,S_n}\keti_{B}^\sigma$
and $\|\alpha\p\keti=\|{\scriptstyle L\p,M\p,T\p,S\p_1,\ldots,S\p_n}\keti_{B}^\sigma$ 
the result is
\beqn\label{nevenspectrum}
&&\tr_{\HH_{\alpha\p \alpha}^{\rm bos}}\left(g^{t}q^{L_0-{c\over 24}}\right)={}_{-t}\brai\alpha\p\|
q^{{1\over 2}(L_0+\ol L_0)-{c\over 24}}\|\alpha\keti_{-t}\\
&&\qquad\qquad=
\sum_{[l_i,m_i,s_i]\in\II_k}
e^{-{2\pi i\over k+2}t(T-T\p+\sum_{i\,{\rm even}}m_i)}
\prod_{i=1}^n\chi_{[l_i,(-1)^{i+1}m_i,s_i]}(q)\nonumber\\
&&\qquad\qquad\qquad\qquad\qquad\times
\prod_{i=2}^n\delta^{(2)}_{s_i,S\p_i-S_i}\N{[L\p,M\p-2T\p,{\scriptstyle \sum_i} S\p_i]}{[L,M-2T,{\scriptstyle
    \sum_i}S_i]}{[l_1,m_1,s_1]*\ldots*[l_n,m_n,s_n]}\,.
\nonumber
\eeqn
As in the case of odd $n$, the shift by 2 of an odd number of $S$-labels 
in one of the boundary states produces the corresponding fermionic spectrum.

The open string spectrum for $\sigma$-permutation boundary conditions 
$\|\alpha\keti=\|{\scriptstyle L,M,T,S_1,\ldots,S_n}\keti_{B}^\sigma$ 
at one end and and tensor product boundary conditions 
$\|\alpha\p\keti=\|{\scriptstyle L\p_1,\ldots,L\p_n,S\p_1,\ldots,S\p_n,M\p={\sum_i} M\p_i}\keti_{B}^\id$
at the other can be calculated to be
\beqn\label{sigmaidneven}
&&\tr_{\HH_{\alpha\p \alpha}^{\rm bos}}\left(g^{t}q^{L_0-{c\over 24}}\right)={}_{-t}\brai\alpha\p\|
q^{{1\over 2}(L_0+\ol L_0)-{c\over 24}}\|\alpha\keti_{-t}\\
&&\qquad\qquad=(k+2)^{n-2\over 2}\sum_{[l,m,s]\in\II_k}\sum_{a\in\ZZ_{k+2}}
e^{{2\pi i\over k+2}t(a-T)}\chi_{[l,m,s]}\left(q^{1\over n}\right)\nonumber\\
&&\qquad\qquad\qquad\qquad\times
\N
{[L\p_1,M\p_1,S\p_1]*\ldots*[L\p_n,M\p_1,S\p_1]}{[L,M-2T+2a+(1-n),{\sum_i}S_i+(1-n)]}{[l,m,s]}
\,.\nonumber
\eeqn
As for the case of odd $n$, the phase in \eq{twcharacter} affects the 
open spectra by a shift in representations $[l,m,s]\mapsto [l,m+(1-n),s+(1-n)]$.

\subsection{Topological spectra}\label{sectopspec}

The topological open spectra associated to the permutation boundary 
conditions described in Section 3.2 can be read off from the full CFT
spectra, by extracting the chiral primary contributions.
In a single minimal model $\cM_k$, chiral primary fields are given by the
highest weight vectors of the representations $[l,l,0]\in\II_k$, in
tensor products of minimal models by tensor products of those.
Thus, in the situation when the open sectors carry a representation
of the sum of all the $N=2$ algebras of the individual minimal models, the 
topological spectra can easily be extracted. 
Otherwise one has much less control of the representation theory and
the identification of chiral primaries can be quite difficult.
As far as permutation boundary conditions are concerned, this more complicated
situation only occurs in sectors of open strings with different
permutation gluing conditions, \cf eqs.\ (\ref{sigmaidnodd},\ref{sigmaidneven}). These cases will be treated at the end of this section and
we start with the cases where the gluing conditions on both sides are
twisted by the same permutation.

From now on, we will restrict our considerations to 
boundary states where all $S$-labels are even. ($S$ odd 
merely corresponds to the opposite choice of spin structure.) 
For a single minimal model, the boundary
spectra \eq{minmodspectrum} then simplify to 
\beqn\label{sevenspectrum}
&&\tr_{\HH_{[L\p,M\p,S\p],[L,M,S]}^{\rm bos}}\left(g^{t} q^{L_0-{c\over 24}}\right)
=\sum_{[l,m,0]\in\II_k}e^{{\pi i t \over k+2}(m-M+M\p)}
\chi_{[l,m,0]}(q)\\
&&\qquad\qquad\qquad\qquad\qquad\qquad\qquad
\times\left(N_{L\p L}^{\;\;l}\;\delta^{(4)}_{S-S\p,0}
+ (-1)^{t}N_{L\p L}^{\;\;k-l}\;\delta^{(4)}_{S-S\p,2}\right)\,,\nonumber
\eeqn
where
\beqn
N_{L\p L}^{\;\;j}=\left\{ \begin{array}{ll}
   1\, &{\rm if}\ \scriptstyle |L-L\p|\leq j
      \leq\min(L+L\p,2k-L-L\p)\ {\rm and}\ L+L\p+j\in 2\ZZ\\ 
   0\,&{\rm otherwise}\end{array}\right.\nonumber
\eeqn
denotes the $\su(2)_k$-fusion rules.

From \eq{sevenspectrum} the topological open spectra can be easily
read off. 
There are bosonic topological open strings with
boundary conditions corresponding to $\|[L,M,0]\keti_B$ and $\|[L\p,M\p,0]\keti_B$
for every $l\in\{0,\ldots,k\}$ such that $N_{L\p L}^{\;\;l}=1$, \ie for all
$l\in\{|L-L\p|,|L-L\p|+2,\ldots,\min(L+L\p,2k-L-L\p)\}$. Their 
$\ZZ_{k+2}$-charges are given by ${1\over 2}(l-M+M\p)$. Likewise,
there are fermionic topological open strings with these boundary conditions 
for every $l\in\{0,\ldots,k\}$ such that $N_{L\p L}^{\;\;k-l}=1$, \ie for all
$l\in k -\{|L-L\p|,|L-L\p|+2,\ldots,\min(L+L\p,2k-L-L\p)\}$. Their $\ZZ_{k+2}$-charges
are given by ${1\over 2}(l-M+M\p+k+2)$. (As expected, a shift by $2$ in the $M$
or $M\p$ shifts the $\ZZ_{k+2}$-charges by 1.)
Thus, the bosonic and fermionic topological Hilbert spaces are given by
\beqn
\HH^0\left(\|{\scriptstyle [L\p,M\p,0]}\keti_B,\|{\scriptstyle [L,M,0]}\keti_B\right)&\cong&
\bigoplus_{\stackrel{l=|L-L\p|}{l+L+L\p\in
    2\ZZ}}^{\min(L+L\p,2k-L-L\p)} \CC_{{1\over 2}(l-M+M\p)}\,,\\
\HH^1\left(\|{\scriptstyle [L\p,M\p,0]}\keti_B,\|{\scriptstyle [L,M,0]}\keti_B\right)&\cong&
\bigoplus_{\stackrel{l=|L-L\p|}{l+L+L\p\in
    2\ZZ}}^{\min(L+L\p,2k-L-L\p)} \CC_{{1\over 2}(-l-M+M\p-2)}\,,
\eeqn
where the subscript $m$ of $\CC_m$ denotes the respective
$\ZZ_{k+2}$-representation. In particular
\beqn
&&\dim \HH^0\left(\|{\scriptstyle [L\p,M\p,0]}\keti_B,\|{\scriptstyle
    [L,M,0]}\keti_B\right)
=\dim \HH^1\left(\|{\scriptstyle [L\p,M\p,0]}\keti_B,\|{\scriptstyle
    [L,M,0]}\keti_B\right) \nonumber\\
&&\qquad\qquad\qquad\qquad\qquad
= \min(L,L\p,k-L,k-L\p) + 1\ . \nonumber
\eeqn
All this information can be summarised in the bosonic and fermionic 
topological partition functions of a single minimal model 
\beqn
\tr_{\HH^0\left(\|[L\p,M\p,0]\keti_B,\|[L,M,0]\keti_B\right)}(g^{t})
&=&
\sum_{l\in\{0,\ldots,k\}}
N_{L\p L}^{\;l}\;\; e^{{2\pi
    it\over k+2}{1\over 2}\left(l-M+M\p\right)}\,,\\
\tr_{\HH^1\left(\|[L\p,M\p,0]\keti_B,\|[L,M,0]\keti_B\right)}(g^{t})
&=&
\sum_{l\in\{0,\ldots,k\}}
N_{L\p L}^{\;k-l}\;\; e^{{2\pi
    it\over k+2}{1\over 2}\left(l-M+M\p+k+2\right)}\; .
\eeqn
For {\sl tensor product} boundary states
$\|\alpha\keti=\|{\scriptstyle L_1,\ldots,L_n,S_1=0,\ldots,S_n=0,M={\textstyle \sum_i}M_i}\keti_{B}^\id$
and 
$\|\alpha\p\keti=\|{\scriptstyle L\p_1,\ldots,L\p_n,S\p_1=0,\ldots,S\p_n=0,M\p={\textstyle \sum_i}M\p_i}\keti_{B}^\id$
we obtain from \eq{tpspectrum}
\beqn
\HH^b\left(\|\alpha\p\keti,\|\alpha\keti\right)=\bigoplus_{\stackrel{b_1,\ldots,b_n\in\ZZ_2}{b+\sum_i
    b_i\in 2\ZZ}}\!\bigotimes_i \;\HH^{b_i}\left(\|{\scriptstyle [L\p_i,M\p_i,0]}\keti_B,\|{\scriptstyle
    [L_i,M_i,0]}\keti_B\right)\ .
\eeqn
For {\sl odd cycle length permutation} boundary states 
$\|\alpha\keti=\|{\scriptstyle L,M,S_i=0}\keti_{B}^\sigma$
and $\|\alpha\p\keti=\|{\scriptstyle
  L\p,M\p,S\p_i=0}\keti_{B}^\sigma$
the topological partition functions follow from \eq{noddspectrum}
\beqn\label{noddtopb}
\tr_{\HH^0\left(\|\alpha\p\keti,\|\alpha\keti\right)}(g^{t})
&=&
\sum_{l_i\in\{0,\ldots,k\}}
N_{L\p L}^{\;l_1*\ldots *l_n}\;\; e^{{2\pi
    it\over k+2}{1\over 2}\left(\sum_i l_i-M+M\p\right)}\,,\\
\tr_{\HH^1\left(\|\alpha\p\keti,\|\alpha\keti\right)}(g^{t})
&=&
\sum_{l_i\in\{0,\ldots,k\}}
N_{L\p L}^{\;(k-l_1)*\ldots *l_n}\;\; e^{{2\pi
    it\over k+2}{1\over 2}\left(\sum_i l_i-M+M\p+k+2\right)}\,.\label{noddtopf}
\eeqn
For {\sl even cycle length permutation} boundary states
$\|\alpha\keti=\|{\scriptstyle L,M,T,S_i=0}\keti_{B}^\sigma$
and $\|\alpha\p\keti=\|{\scriptstyle
  L\p,M\p,T\p,S\p_i=0}\keti_{B}^\sigma$
they can be extracted from \eq{nevenspectrum} to be
\beqn\label{neventopb}
\tr_{\HH^0\left(\|\alpha\p\keti,\|\alpha\keti\right)}(g^{t})\!\!\!\!
&=&
\!\!\!\!\!\!\!\!\!\!\!\sum_{l_i\in\{0,\ldots,k\}}\!\!\!\!\!\!\!
N_{L\p L}^{\;l_1*\ldots *l_n}\;\delta^{\scriptscriptstyle
  (2k+4)}_{\scriptscriptstyle M-M\p-2(T-T\p),\sum_i(-1)^{i+1}l_i}
e^{\scriptscriptstyle {\pi
    it\over k+2}\left(\sum_{i} l_{i}-M+M\p\right)}\,,\phantom{***aaa}\\
\tr_{\HH^1\left(\|\alpha\p\keti,\|\alpha\keti\right)}(g^{t})
\!\!\!\!&=&\!\!\!
\!\!\!\!\!\!\!\!\sum_{l_i\in\{0,\ldots,k\}}\!\!\!\!\!\!\!\!
N_{L\p L}^{\;(k-l_1)*\ldots *l_n}\,\delta^{\scriptscriptstyle
 (2k+4)}_{\scriptscriptstyle M\!-\!M\p\!-2(T\!-\!T\p),\sum_i(-1)^{i+1}l_i+k+2}
e^{\scriptscriptstyle {\pi
    it\over k+2}\left(\sum_{i} l_{i}-M+M\p\right)}.\label{neventopf}
\eeqn
As alluded to above, 
the extraction of these topological spectra from the corresponding
CFT spectra heavily relied on the fact that chiral primary
states in tensor products of minimal models are tensor products of
minimal model chiral primary states. The Hilbert spaces of open
strings satisfying boundary conditions with {\sl different permutations on
both sides} however do not carry a representation of the tensor product
of the minimal model $N=2$ algebras. Rather, they decompose into
twisted representations of a $\ZZ_n$-orbifold thereof, where 
$\ZZ_n$ is generated by the permutation $\sigma$. 

We can identify the chiral primaries amongst the highest weight vectors of
the twisted representations by their characteristic relation between 
conformal weight $h$ and $U(1)$-charges $q$, namely $h={1\over 2}q$ 
(which holds in unitary theories). Conformal weight $\hat h$ and 
$U(1)$-charge $\hat q$ of the $\ZZ_n$-twisted representations with
character $\chi_{[l,m,s]}(q^{1\over n})$ can be expressed in terms of
the conformal weight $h$ and $U(1)$-charge $q$ of the respective representation
with character $\chi_{[l,m,s]}(q)$ in the ``mother'' theory as (for more
details on cyclic orbifolds see \eg \cite{bhs97})
\beq
\hat h={h\over n}+{c\over 24}\left(n-{1\over n}\right)\,,\qquad
\hat q=q\,.
\eeq
The chiral primary condition $\hat h={1\over 2}\hat q$ can therefore
be expressed in terms of $h$ and $q$ as follows
\beq
h+{1-n\over 2}q+{c\over 6}\left({1-n\over 2}\right)^2={1\over
  2}\left(q+{c\over 3}\left({1-n\over 2}\right)\right)\,.
\eeq
This, however, is nothing else than the chiral primary condition for the
representation obtained from the original one after spectral flow ${\cal
  U}_\eta$ by $\eta={1-n\over 2}$ units. Namely, conformal weight and 
$U(1)$-charge change under the spectral flow ${\cal U}_\eta$ as
\beq
h\mapsto h_\eta=h+\eta q+{c\over 6}\eta^2\,,\qquad 
q\mapsto q_\eta=q+{c\over 3}\eta\,.
\eeq
The action of this spectral flow on representations
is given by 
\beq\label{identificationshift}
{\cal U}_{1-n\over 2}[l,m,s]=[l,m-(1-n),s-(1-n)]\,.
\eeq
Therefore, a representation with twisted character
$\chi_{[l,m,s]}(q^{1\over n})$ is built on a 
chiral primary highest weight state iff the minimal model representation
${\scriptstyle[l,m-(1-n),s-(1-n)]}$ is built on a chiral primary.

Having managed to identify the chiral primaries in the twisted
representations, it is not difficult to extract the topological partition
functions between permutation boundary states 
$\|\alpha\keti=\|{\scriptstyle L,M,S_i=0}\keti_{B}^\sigma$
(for $n$ odd) or $\|\alpha\keti=\|{\scriptstyle L,M,T,S_i=0}\keti_{B}^\sigma$
(for $n$ even) and a tensor production boundary state 
$\|\alpha\p\keti=\|{\scriptstyle L\p_1,\ldots,L\p_n,S\p_i=0,M\p={\sum_i} M\p_i}\keti_{B}^\id$
from the CFT-spectra\footnote{
Note that the shift \eq{identificationshift}, which is used to identify
chiral primaries among the twisted open CFT states, is exactly opposite
to the shift in the open spectra produced by the relative phases
between twisted characters and $\sigma$-twisted traces, \cf the end 
of Section 3.2.}  
\eq{sigmaidnodd}
\beqn\label{sigmaidtopb}
\tr_{\HH^0\left(\|\alpha\p\keti,\|\alpha\keti\right)}(g^{t})
&=&{ (k+2)^{\left[{n-1\over 2}\right]}}\!\!\!\!
\sum_{l\in\{0,\ldots,k\}}\!\!\!\!
N_{L\p_1*\ldots*L\p_n L}^{\;l}\;\; e^{{2\pi
    it\over k+2}{1\over 2}\left(l-M+M\p\right)}\,,\\
\tr_{\HH^1\left(\|\alpha\p\keti,\|\alpha\keti\right)}(g^{t})
&=&{ (k+2)^{\left[{n-1\over 2}\right]}}\!\!\!\!
\sum_{l\in\{0,\ldots,k\}}\!\!\!\!
N_{L\p_1*\ldots*L\p_n L}^{\;k-l}\;\; e^{{2\pi
    it\over k+2}{1\over 2}\left(l-M+M\p+k+2\right)}\,,\phantom{***}\label{sigmaidtopf}
\eeqn
where $[\,\cdot\,]$ denotes the integer part.
\subsection{Boundary states in Gepner models}\label{sect:bdrygepner}
In this section, we would like to recall briefly how to extract
information about Gepner model branes from the boundary states 
in tensor products of minimal models discussed above. 

Gepner models consist of orbifolds of tensor products \eq{tpmodel}
of $N=2$ minimal models coupled to some free external theory, where 
the orbifold construction implements the GSO-projection of the
internal part.
The orbifold group is the cyclic group $\Gamma=\ZZ_{H}$, generated by 
the product of the generators of the $\ZZ_{k_i+2}$-symmetry groups of 
the individual minimal models. Hence, $H={\rm lcm}(k_1+2,\ldots,k_n+2)$.  
If the ``Calabi-Yau condition''$\sum_{i=1}^n(k_i+2)^{-1}=1$ is satisfied, 
then this model describes a string compactification on the hypersurface 
in weighted projective space defined by the vanishing of the superpotential 
$W=x_1^{k_1+2}+\ldots+x_n^{k_n+2}$.

Above, we constructed certain boundary states in tensor products of 
minimal models, and there is a standard procedure of obtaining boundary 
conditions in orbifold models from those of the original unorbifolded 
theories (see the remarks in Section 3.1): Starting from a 
boundary state in the original model which is invariant under the action 
of the orbifold group $\Gamma$, one sums up all the states obtained from 
it by twisting with elements of $\Gamma$, then divides by $\sqrt{|\Gamma|}$ 
to ensure correct normalisation. Obviously this has the effect of projecting 
the corresponding open string sectors to the trivial representations of 
the orbifold group. 

In the preceding sections, we obtained such $\Gamma$-invariant
boundary states in tensor products of minimal models, and we also presented
all the twisted boundary states. Summing up all these twisted components,
one arrives at the internal parts of the respective boundary states
in Gepner models. From the $\Gamma$-twisted open partition functions 
in the tensor products of minimal models determined in Section 
\ref{sectopspec}, one can read off the respective open sectors of 
the (internal part of the) Gepner model, simply by extracting the 
$\Gamma$-invariant parts.

Boundary states in full Gepner models can be obtained as tensor product of 
boundary states of the internal and the external theories respectively. 
However, the alignment of NS- and R-sectors has to be ensured in this construction, intertwining the two factor states in a non-trivial 
way. Nevertheless, 
certain ``invariants'' of boundary conditions, which only depend on the 
RR- and the NSNS-part of the boundary states separately, factorise into 
internal and external contributions. This is true in particular for the 
open string Witten index 
\beqn
I(\alpha\p,\alpha)&=&
{}_{RR}\brai\alpha\p\|(-1)^{F_L}q^{{1\over 2}(L_0+\ol L_0)-{c\over 24}}\|\alpha\keti_{RR}\nonumber\\
&=&
\dim\HH^0\left(\|\alpha\p\keti,\|\alpha\keti\right)
-\dim\HH^1\left(\|\alpha\p\keti,\|\alpha\keti\right)\,,\nonumber
\eeqn
where $F_L$ is the holomorphic fermion number on the bulk Hilbert space
(see \eg \cite{Brunner:1999jq,hiv}).
Therefore it makes sense to calculate $I$ for the internal part
of the Gepner model boundary state alone, \ie in the orbifold of the
tensor products of minimal models. 

This index can be calculated easily from the topological open 
partition functions for tensor product bulk theories. One merely 
needs to identify the respective $\Gamma$-invariant parts of the bosonic 
and fermionic topological Hilbert spaces and subtract their dimensions.

For example, the Witten index between the tensor product boundary 
state 
$\|\alpha\p\keti=\|{\scriptstyle L\p_i=0,S\p_i=0,M\p}\keti_{B}^\id$
and a permutation brane
$\|\alpha\keti=\|{\scriptstyle L,M,S_i=0}\keti_{B}^\sigma$ for odd $n$ or
$\|\alpha\keti=\|{\scriptstyle L,M,T,S_i=0}\keti_{B}^\sigma$ for even $n$  
can be obtained by summing over $t$ in 
eqs.~(\ref{sigmaidtopb},\ref{sigmaidtopf}) and dividing by $|\Gamma|=k+2$
(here assuming $k_i = k$ and also $n= k+2$ for notational simplicity). 
In this way we arrive at
\[
I(\alpha\p,\alpha) = (k+2)^{\left[\frac{n-1}{2}\right]}\; 
  \left( \delta^{(2k+4)}_{L-M+M',0} - \delta^{(2k+4)}_{L+M-M',2k+2} \right),
\]
Defining the parameters $\mu(L,M):={1\over 2}(L-M)\in\ZZ_{k+2}$, and 
extending them additively to tensor product boundary conditions, this 
expression can be written in terms of a $(k+2)\times(k+2)$ shift matrix
$G_{\mu\p\mu}=\delta^{(k+2)}_{\mu-\mu\p+1,0}$ as
\begin{equation}\label{eq:wpiid}
  I^{\id\,\sigma}(L\p_i=0,L)_{\mu\p\mu} 
    = (k+2)^{[\frac{n-1}{2}]} ( 1 - G^{-L-1} )_{\mu\p\mu}.
\end{equation}
This can of course be generalised to permutations consisting of $N$ cycles 
of length $n_\nu$ and labels $L_\nu$. The Witten indices for open strings 
between such a brane and a $L\p_i =0$ tensor product brane are encoded in 
the matrix
\begin{equation}\label{iidsigma}
  I^{\id \sigma} = \prod_{\nu=1}^N\ (k+2)^{\left[\frac{n_\nu-1}{2}\right]}\; 
  (1 - G^{-L_\nu-1})\ .
\end{equation}
All the Witten indices can be written in terms of $G$, but the expressions 
for $I^{\id\,\sigma}$ with arbitrary $L\p_i$ and those for $I^{\sigma\,\sigma}$ appear to be more involved than \eq{eq:wpiid}; see \cite{r02} for some results 
in the quintic case. Nevertheless, even in the absence of a closed formula one 
can extract each index in a straightforward manner from the topological 
partition functions determined in Section \ref{sectopspec}.
In Table \ref{inttablepp} we list some $I^{\sigma\,\sigma}$ for $k=1,2,3$
and various values of $n$, $L\p$, $L$, $m=T+{1\over 2}(L-M)$ 
and $\Delta m=m-m\p$ (the latter being defined for even $n$ only),  
and in Table \ref{inttableip} some $I^{\id\,\sigma}$ for $k=1,2,3$ and 
various $n$, $L\p_i$ and $L$.

\begin{table}
\scriptsize
\centering
\begin{tabular}{cc}
\begin{minipage}{2.65 in}
\begin{tabular}{|@{ }>{$}c<{$}@{ }|@{ }>{$}c<{$}@{ }|@{ }>{$}c<{$}@{ }>{$}c<{$}@{\,}>{$}c<{$}@{}|@{ }>{$}c<{$}@{ }|}
\hline
 k & n & L\p & L & \Delta m & I^{\sigma\sigma}(\alpha\p,\alpha)\\
\hline
1 & 3 & 0 & 0 &  & -3 G^{2}+3 G\\
\hline
2 & 3 & 0 & 0 &  & -6 G^{3}+6 G^{2}+2 G-2\\
 & 3 & 0 & 1 &  & -8 G^{3}+8 G\\
 & 3 & 1 & 1 &  & -8 G^{3}+8 G^{2}+8 G-8\\
 & 4 & 0 & 0 & 0 & 4 G^{3}+10 G^{2}+4 G+{2}\\
 & 4 & 0 & 0 & 1 & -2 G^{3}+4 G^{2}-2 G-4\\
 & 4 & 0 & 0 & 2 & -4 G^{3}+2 G^{2}-4 G-6\\
 & 4 & 0 & 0 & 3 & -2 G^{3}+4 G^{2}-2 G-4\\
 & 4 & 0 & 1 & 0 & 8 G^{2}+8 G\\
 & 4 & 0 & 1 & 1 & -8 G^{3}-8\\
 & 4 & 0 & 1 & 2 & -8 G^{3}-8\\
 & 4 & 0 & 1 & 3 & 8 G^{2}+8 G\\
 & 4 & 1 & 1 & 0 & 8 G^{3}+16 G^{2}+8 G\\
 & 4 & 1 & 1 & 1 & 8 G^{2}-8\\
 & 4 & 1 & 1 & 2 & -8 G^{3}-8 G-16\\
 & 4 & 1 & 1 & 3 & 8 G^{2}-8\\
\hline
3 & 3 & 0 & 0 &  & -10 G^{4}+10 G^{3}+5 G^{2}-5\\
 & 3 & 0 & 1 &  & -15 G^{4}+15 G^{2}+5 G-5\\
 & 3 & 1 & 1 &  & -15 G^{4}+15 G^{3}+20 G^{2}-20\\
 & 4 & 0 & 0 & 0 & 10 G^{4}+20 G^{3}+10 G^{2}+5 G+{5}\\
 & 4 & 0 & 0 & 1 & -5 G^{4}+5 G^{3}-5 G^{2}-10 G-10\\
 & 4 & 0 & 0 & 2 & -5 G^{4}+5 G^{3}-5 G^{2}-10 G-10\\
 & 4 & 0 & 0 & 3 & 10 G^{3}-5 G-5\\
 & 5 & 0 & 0 &  & 125 G^{4}+125 G^{3}-125 G^{2}-125 G\\
 & 5 & 0 & 1 &  & 125 G^{4}+250 G^{3}-250 G-125\\
 & 5 & 1 & 1 &  & 375 G^{4}+250 G^{3}-250 G^{2}-375 G\\
\hline
\end{tabular}
\caption{Witten index $I^{\sigma\sigma}$}
\label{inttablepp}
\end{minipage}
&
\begin{minipage}{2.65 in}
\begin{tabular}{|@{ }>{$}c<{$}@{ }|@{ }>{$}c<{$}@{ }>{$}c<{$}@{ }|@{ }>{$}c<{$}@{ }|}
\hline
 k & (L_i') & L & I^{\id\,\sigma}(\alpha\p,\alpha)\\
\hline
1 & (0, 0, 0) & 0 & -3 G^{2}+{3}\\
\hline
2 & (0, 0, 0) & 0 & -4 G^{3}+{4}\\
 & (0, 0, 1) & 0 & -4 G^{3}+4 G\\
 & (0, 1, 1) & 0 & -4 G^{3}+4 G^{2}+4 G-4\\
 & (1, 1, 1) & 0 & 8 G^{2}-8\\
 & (0, 0, 0) & 1 & -4 G^{2}+{4}\\
 & (0, 0, 1) & 1 & -4 G^{3}-4 G^{2}+4 G+{4}\\
 & (0, 1, 1) & 1 & -8 G^{3}+8 G\\
 & (1, 1, 1) & 1 & -8 G^{3}+8 G^{2}+8 G-8\\
 & (0, 0, 0, 0) & 0 & -4 G^{3}+{4}\\
 & (0, 0, 0, 1) & 0 & -4 G^{3}+4 G\\
 & (0, 0, 1, 1) & 0 & -4 G^{3}+4 G^{2}+4 G-4\\
 & (0, 1, 1, 1) & 0 & 8 G^{2}-8\\
 & (1, 1, 1, 1) & 0 & 8 G^{3}+8 G^{2}-8 G-8\\
\hline
3 & (0, 0, 0) & 0 & -5 G^{4}+{5}\\
 & (0, 0, 1) & 0 & -5 G^{4}+5 G\\
 & (0, 1, 1) & 0 & -5 G^{4}+5 G^{2}+5 G-5\\
 & (1, 1, 1) & 0 & -5 G^{4}+5 G^{3}+10 G^{2}-10\\
 & (0, 0, 0) & 1 & -5 G^{3}+{5}\\
 & (0, 0, 1) & 1 & -5 G^{4}-5 G^{3}+5 G+{5}\\
 & (0, 1, 1) & 1 & -10 G^{4}-5 G^{3}+5 G^{2}+10 G\\
 & (1, 1, 1) & 1 & -15 G^{4}+15 G^{2}+10 G-10\\
 & (0, 0, 0, 0) & 0 & -5 G^{4}+{5}\\
 & (0, 0, 0, 0, 0) & 0 & -25 G^{4}+{25}\\
 & (0, 0, 0, 0, 1) & 0 & -25 G^{4}+25 G\\
 & (0, 0, 0, 1, 1) & 0 & -25 G^{4}+25 G^{2}+25 G-25\\
\hline
\end{tabular}
\caption{Witten index $I^{\id\,\sigma}$}
\label{inttableip}
\end{minipage}\\
\end{tabular}
\end{table}

\section{Permutation branes and linear matrix factorisations}
 
A tensor product $\cM_k^{\otimes n}$ of $N=2$ minimal models describes
the critical behaviour of a Landau-Ginzburg model with superpotential 
$W=x_1^d+\ldots+x_n^d$ for $d=k+2$, on a world-sheet without or with
boundary. Therefore, one can expect that the CFT B-type branes from 
above have some counterpart in the form of an LG boundary condition -- 
more concretely that the topological information of the CFT brane 
can be encoded in a matrix factorisation of the LG potential $W$. 
Our proposal is that topological permutation branes correspond to 
certain {\sl linear matrix factorisations}.

\subsection{Linear matrix factorisations}

A linear matrix factorisation \cite{Backelin:1988} of a homogeneous polynomial 
$W$ of degree $d$ in the variables $x_1,\ldots,x_n$ is given by a set of $d$ square
matrices $\alpha_0,\ldots,\alpha_{d-1}$ over $\CC[x_1,\ldots,x_n]$ all
of which are linear in the $x_i$ and satisfy 
\begin{equation}
  \alpha_0 \,\alpha_1 \cdots \alpha_{d-1} = W\; {\bf 1} \ \ .
\end{equation}
From the $\alpha_i$, we can obtain two-factor matrix factorisations by 
choosing
\beq\label{ordfromlin}
p_0=\alpha_{\pi(0)}\cdots\alpha_{\pi(\ell -1)}\;\;\;{\rm and}\;\;\;
p_1=\alpha_{\pi(\ell)}\cdots\alpha_{\pi(d-1)}
\eeq
for $0<\ell<d-1$ and any cyclic permutation $\pi$ of
$(0,\ldots,d-1)$.

A special class of linear matrix factorisations of $W = \sum x_i^d$ have been constructed  explicitly 
by Backelin, Herzog and Sanders in \cite{Backelin:1988}.  For
all homogeneous polynomials there exists a unique (up to equivalence and
cyclic permutation of the factors) indecomposable linear matrix
factorisation with the property
\begin{equation}
  \alpha_t(x_i) \alpha_{t+1}(x_j) = \xi \alpha_t(x_j) \alpha_{t+1}(x_i) \qquad i>j
\end{equation}
where $\alpha_t(x_i)$ is the matrix obtained from $\alpha_t(x_1,\ldots,x_n)$ 
by setting $x_j=0$ for $j \ne i$, and $\xi$ is a primitive $d$th root of unity.

In the case $W = x_1^d + \ldots x_n^d$, these factorisations consist of $d^\gamma
\times d^\gamma$ matrices, $\gamma= \left[\frac{n-1}{2}\right]$, which can be written as 
\begin{equation}\label{alphadef}
  \alpha_i = x_1 + \xi^i\; \alpha_{d,n}, 
\end{equation}
where 
the $\alpha_{d,n}$ can be defined by a recursion formula as follows:
One introduces $d\times d$ matrices 
\begin{equation}\label{epsilons}
  (\epsilon_1)_{ij} = \xi^{i-1}\; \delta_{i,j-1}\ , \quad
   (\epsilon_2)_{ij} = 
\xi^{i-1}\;\delta_{i,j}\,
, \quad
  (\epsilon_3)_{ij} = \delta_{i,j-1}\ , 
\end{equation}
where all Kronecker deltas are understood modulo $d$, as well as the number 
\beq
\mu_n=\begin{cases}
1\;& d\ {\rm odd} \\
\eta\;& d\ {\rm even}\ {\rm and}\ \left[{n-1\over 2}\right] \ {\rm even}\\
\eta^{-1}\;&d\ {\rm even} \ {\rm and} \ \left[{n-1\over 2}\right] \ {\rm odd}\\
\end{cases}\,,
\eeq
$\eta$ being a primitive $d$th root of $-1$ with $\eta^2=\xi$. Using these, 
one defines
\begin{align}
  \alpha_{d,1}&=0, \qquad \alpha_{d,2}=\mu_2\, x_2, \\
  \alpha_{d,n+2} &= \epsilon_2 \otimes \alpha_{d,n}
                   + \epsilon_3 \otimes \mu_{n+2}\,x_{n+1}\,{\bf 1} + \epsilon_1 \otimes x_{n+2}\,{\bf 1},
\label{recformula}
\end{align}
where the {\bf 1}'s stand for identity matrices of the same size as $\alpha_{d,n}$.  

These special linear matrix factorisations have certain nice properties. 
They are homogeneous in the $x_i$ and indecomposable (\ie not equivalent 
to direct sums). Moreover, it is obvious from \eq{alphadef} that all 
the $\alpha_i$ commute, which in particular means that they give rise 
to matrix factorisations \eq{ordfromlin} not only for $\pi$ cyclic but 
for all permutations $\pi\in S_d$. Thus, for every proper subset 
$I\subset\{0,\ldots,d-1\}$ we obtain the two-factor matrix factorisations
\beq\label{defmf}
M_{I,I^c}=\left(p_0=\prod_{i\in I}\alpha_i,\  p_1=\prod_{i\in I^c}\alpha_i\right)\,,
\eeq
where $I^c=\{0,\ldots,d-1\}\backslash I$. For every $\ell=|I|$ 
these are ${d\choose\ell}$ ones.

Note, however, that not all of them have to be inequivalent. 
To determine, for given number of variables $n$, the possible
equivalences 
(as defined in \eq{equivMF}) between them, first note that since the $p_i$ above are 
homogeneous, we can restrict to constant matrices $U_n,V_n$ in 
eq.\ \eq{equivMF}, and that two matrix factorisations associated 
to index sets $I$ and $I'$ as above can be equivalent only if they are
of the same degree, \ie if $|I|=|I'|$. 
The specific form $p_0 = x_1^{|I|} + \ldots$ enforces $U_n=V_n$, 
and exploiting \eq{alphadef} further one finds that 
$U_n^{-1}\alpha_{d,n}U_n=\xi^i \alpha_{d,n}$ has to hold 
for some integer $i$. Conjugation of a matrix factorisation \eq{defmf} 
with such a $U_n$ then just shifts the set $I$ to $I+i$ (understood 
modulo $d$). 

To proceed, one observes that given $U_n$ for a fixed $n$, 
one obtains a matrix $U_{n+2}$ conjugating $\alpha_{d,n+2}$ to 
$\xi^i\alpha_{d,n+2}$ by setting $U_{n+2}=\epsilon_2^i\otimes U_n$. 
Vice versa, using the explicit form of the matrices $\epsilon_m$ and 
inspecting the recursion relation \eq{recformula} for $\alpha_{d,n+2}$, one can show 
that any $U_{n+2}$ with the correct conjugation property can be formed 
from a $U_n$ in this way. 

This allows us to list the classes of inequivalent matrix factorisations of 
the type \eq{defmf}:  For {\sl odd} $n>1$, one constructs possible
equivalences $U_n$ starting from $U_1=1$, which obviously conjugates
$\alpha_{d,1}=0$ to $\xi^i\alpha_{d,1}$. Therefore, in this case
matrix factorisations \eq{defmf} defined 
by the sets $I$ and $I\p$ are equivalent if and only if $I\p$ is a shift of $I$. 
On the other hand, since $\alpha_{d,2}$ is a non-zero rank-1 matrix, there 
is no matrix $U_2$ to non-trivially conjugate it. Therefore, for {\sl even} $n$ 
all the factorisations \eq{defmf} are inequivalent.

\vspace{0.1cm}

The case $n=1$ provides the simplest example of linear factorisations,
where  $\alpha_i=x_1$ for all $i$ and we obtain the 
well-known $d-2$ inequivalent factorisations of minimal model 
potential $W=x_1^d$, by grouping together $\ell$ factors $x_1$ 
into $p_0$ and the remaining $d-\ell$ into $p_1$. For simplicity, 
they will be denoted $M_\ell(x_1)$ in the following. 

The next example $n=2$ is a little more interesting. Here, the linear 
matrices are given by 
$\alpha_i=x_1+\mu_2\xi^ix_2$, so that we obtain a matrix factorisation 
with $p_0=\prod_{i\in I}(x_1+\mu_2\xi^i x_2)$ and $p_1=\prod_{i\in I^c}(x_1+\mu_2\xi^i x_2)$
for every proper subset $I\subset\{0,\ldots,d-1\}$. These factorisations 
were introduced into the discussion of B-branes in LG models in
\cite{addf04} and then related to CFT permutation branes with 
$\sigma = (1\,2)$ in \cite{Brunner:2005fv}\footnote{Note however, that
  our factorisations differ from the ones used in
  \cite{addf04,Brunner:2005fv} by a shift $x_2\mapsto
  \xi^{\left[{d+1\over 2}\right]}x_2$.}. 

For $n>2$, the factorisations are much harder to treat `by hand' since
the size $d^{[{n-1\over 2}]}$ of the matrices grows exponentially with
$n$, which is why later on we will partly resort to computer algebra
programmes to perform some of the computations. Note that 
factorisations \eq{defmf} for the case $n=3$ and $d=3$ have already occurred 
in the classification of maximal Cohen-Macaulay modules over the
cone of the elliptic curve in \cite{Laza:2002,Kahn:1989} and 
in the discussion of D-branes on the elliptic curve 
in \cite{Hori:2004ja,Brunner:2004mt}. 

Ultimately, we are interested in graded matrix factorisations
\beq
(P_1,\rho_1)\overset{p_1}{\underset{p_0}{\rightleftarrows}} (P_0,\rho_0)\,,
\eeq
where apart from
the matrix factorisation itself, $\ZZ_d$-representations $\rho_i$ on the $P_i$ 
are specified, which are compatible with the module structure (recall that $R$ 
is graded) and the maps $p_i$. 

For indecomposable matrix factorisations as in \eq{defmf}, there is only a
choice of one irreducible representation $\alpha$ of $\ZZ_d$, which determines 
$\rho_1,\rho_0$ completely, and we specify it by setting the 
degree of the element $1\in R\subset P_0=R^{d^\gamma}$ to be 
$\alpha\in\ZZ_{k+2}$. 
We denote the corresponding graded factorisations $M^\alpha$. 

\subsection{Relation to permutation branes}

We would now like to compare these linear matrix factorisations to 
the boundary states constructed in the previous sections.
This will be done by analysing the open topological string sectors 
on the matrix factorisation side, \ie the graded Ext-groups
$\mathrm{Ext}(P,Q)$ between modules $P=\coker p_1$, $Q=\coker q_1$
corresponding to matrix factorisations $(p_0,p_1)$, $(q_0,q_1)$, and
comparing them to the respective CFT-results obtained in Section
\ref{sectopspec}. 
Here, we will consider the cases where these matrix factorisations 
are linear factorisations in $n$ variables or tensor 
products $M_{L_1,\ldots,L_n}^{\otimes}:=M_{L_1}(x_1)\otimes\ldots
\otimes M_{L_n}(x_n)$ of linear matrix factorisations in one variable,
\cf \eq{tpfactorisation}.  The 
generalisation to tensor products of multi-variable linear 
matrix factorisations is straightforward. 

The $\ZZ_d$-representation of the linear factorisations
$M_{L}(x)$ is specified by $\alpha=\rho_0$, and the 
$\ZZ_d$-representations of 
$M_{L_1}^{\alpha_1}(x_1)\otimes\ldots\otimes M_{L_n}^{\alpha_n}(x_n)$ 
only depends on $\sum_i\alpha_i$. We define
$M_{L_1,\ldots,L_n}^{\otimes\,\alpha}$ 
to be this tensor product factorisation for an arbitrary partition 
$\alpha=\sum_i\alpha_i$.
These tensor product matrix factorisations reproduce the topological spectra of 
tensor product boundary states \eq{tpbd} (a discussion of this can 
be found in \cite{Ashok:2004zb}), the precise correspondence being
\beq\label{correspondencetp}
\|L_1,\ldots,L_n,S_1,\ldots,S_n,M={\textstyle \sum_i} L_i-2\alpha\keti_{B}^\id 
\quad\longmapsto\quad  M_{L_1,\ldots,L_n}^{\otimes\,\alpha}\, .
\eeq
We propose the following {\sl correspondence between CFT permutation 
boundary states and matrix factorisations}:
\beq\label{correspondence}
\begin{array}{lccc}
n\;{\rm odd}:&\|{\scriptstyle L,L-2\alpha,S_1=0,\ldots,S_n=0}\keti_B^\sigma&
 \longmapsto& M_{\{0,\ldots,L\},\{L+1,\ldots,d-1\}}^\alpha \,,\\
n\;{\rm even}:& \|{\scriptstyle L,L-2\alpha,T=m-\alpha,S_1=0,\ldots,S_n=0}\keti_B^\sigma
& \longmapsto& M_{\{0,\ldots,L\}-m,\{L+1,\ldots,d-1\}-m}^\alpha \,,
\end{array}
\eeq
where we use the notations of  \eq{defmf}, and where elements in the
sets $I$ are understood to be taken modulo $d=k+2$.

Note that, for odd $n$, factorisations $M_{I,I^c}$ and $M_{I+i,I^c+i}$
are equivalent, whereas for even $n$ this is not the case and the
respective shift $m$ is determined by the boundary state labels
$(L,M,T)$ through $m=T+{1\over 2}(L-M)$.

\vspace{.2cm}

For the case $n=1$, a single minimal model, this correspondence is of 
course well known. We have spelled out the topological spectra in 
Section \ref{sectopspec}, and the Ext-groups of the corresponding 
matrix factorisations can easily be calculated, see \eg 
\cite{kl03b,h04,Ashok:2004zb}.

For the next complicated case $n=2$, the linear matrix factorisations 
still have rank one, so the correspondence can still be checked by hand 
in a straightforward way. Ext-groups involving two $n=2$ linear factorisations
or one linear and one tensor product factorisation were first studied
in \cite{Ashok:2004zb}. The comparison with CFT permutation boundary 
states for $\sigma = (1\,2)$ has been carried out in great detail in the 
recent work \cite{Brunner:2005fv}, so we refrain from repeating the 
calculations here.

Whenever $n>2$, the linear matrix factorisations involve higher rank 
matrices, and the computation of the $\Ext$-groups may become quite 
tedious. We do not yet have a general 
derivation for all possible combinations of linear and tensor 
product factorisations. The 
$\Ext$-groups between $M_{I,I^c}^\alpha$ and 
$M_{L_1,L_2=0,\ldots,L_n=0}^{\otimes,\beta}$ are calculated for 
arbitrary $n$ and $d$ in Section \ref{evidence}, exploiting certain 
constructions from homological algebra. The results  
are in agreement with the correspondence proposed above. 

To check agreement also for the other spectra (in particular the ones involving 
two higher rank linear matrix factorisations), we resort to 
calculating the respective $\Ext$-groups on a case-by-case basis on
the computer.
For this purpose we used the computer algebra 
program Macaulay2 \cite{M2}. Some of the results of these calculations
are presented in Section \ref{comtests} below. All tests show agreement 
with the CFT results obtained in Chapter \ref{cftsection} and confirm 
the correspondence \eq{correspondence}. 

\vspace{.1cm}

Before we turn to Ext-groups, we can apply a simpler test to our correspondence between linear 
matrix factorisations and boundary states, concerning the behaviour under 
the charge symmetry $\ZZ_d^n$, whose generators act as $g_i\,:\ x_j \mapsto 
\xi^{\delta_{ij}}\,x_j$ on the LG variables and multiply CFT Ishibashi states 
by $\xi^{{1\over 2}(m_i+\ol{m}_i)}$. From formulae (\ref{noddbd},\ref{nevenbd})  for  the permutation 
boundary states,  one sees that each $g_i$ shifts the $T$-label by
$(-1)^{i}$ for $n$ even, 
while it leaves the boundary state labels invariant when $n$ is odd. For the linear 
matrix factorisations, on the other hand, $g_i$ induces a shift 
$I \mapsto I + (-1)^{i}$ of the index set -- which is an equivalence
for $n$ odd, but changes the equivalence class of the matrix
factorisation for even $n$, in accordance with the proposed correspondence.

\subsection{Calculation of some $\Ext$-groups}\label{evidence}

Let $M_{L_1,\ldots,L_n}^\otimes=(q_0,q_1)$ be a tensor product matrix 
factorisation as above and $M_{I,I^c}=(p_0,p_1)$ with $|I|=L+1$ any linear 
matrix factorisation of degree $L+1$. In this section, we aim at calculating 
the groups $\Ext_R(\coker q_1 ,\coker p_1 )$. To do this, we use a relation 
between the factorisation $(q_0,q_1)$ and the module 
$N=N_{L_1,\ldots,L_n}:=R/(x_1^{L_1+1},\ldots,x_n^{L_n+1})$.
One way to establish this connection, namely by deconstructing  the 
tensor product matrix factorisation $(q_0,q_1)$, is presented in 
Appendix \ref{tpfactapp}. Here, we will take a more direct approach and 
construct a free resolution of $N$ which becomes $2$-periodic after 
$(n-1)$ steps with periodic part given by $(q_0,q_1)$.
In fact, this is a special case of a more general construction due to Eisenbud
\cite{Eisenbud:1980}. For a commutative ring $A$ and an ideal ${\mathcal I}$,
Eisenbud constructs a free resolution of a $B=A/{\mathcal I}$-module $V$ out of
an $A$-free resolution of $V$.

In our case $A=\CC[x_1,\ldots,x_n]$, ${\mathcal I}=(W=\sum_i x_i^d)$ and $B=R$. As $A$-free
resolution of $N$ we use the {\sl Koszul complex} of $(x_1^{L_1+1},\ldots,x_n^{L_n+1})$
which is a minimal $A$-free resolution of $N$ of length $(n+1)$
\beq
0\longrightarrow K_n\stackrel{\delta}{\longrightarrow} K_{n-1} \stackrel{\delta}{\longrightarrow} \ldots
\stackrel{\delta}{\longrightarrow} K_1\stackrel{\delta}{\longrightarrow} K_0\longrightarrow N\longrightarrow 0
\,,
\eeq
where $K_i=\Lambda^i V$ is the $i^{\rm th}$ exterior power of 
$V=A^n$ with $A$-basis $\{e_1,\ldots,e_n\}$ and co-differential 
\beq\label{differentialdeltadef} 
\delta\, e_{i_1} \we \cdots \we e_{i_p}  := 
    \sum_{j=1}^p  (-1)^{j-1}  x_{i_j}^{L_{i_j}+1}
     e_{i_1} \we \cdots \we e_{i_{j-1}} \we e_{i_{j+1}} \cdots \we e_{i_p}\,.
\eeq
To obtain an $R$-free resolution of $N $ from this, one first 
introduces the maps $\sigma:K_i\longrightarrow K_{i+1}$ defined by
\beq 
\sigma: \omega \longmapsto \left(\sum_i x_i^{d-L_{i}-1} e_i
\right)\we \omega\ ,
\eeq
which satisfy $\delta\sigma+\sigma\delta=W$.
Furthermore let $T_i :=A\, t^i$ for $i\geq 0$ and define the operator 
$\lambda:T_n\longrightarrow T_{n-1}$ by 
$\lambda(t^n)=t^{n-1}$ for $n\geq 1$ (and $\lambda:=0$ on $T_0$). 
Then we obtain the chain of $A$-modules 
\beq\label{Fcomplex}
\ldots\longrightarrow F_l\stackrel{\wt\delta}{\longrightarrow} F_{l-1} 
\stackrel{\wt\delta}{\longrightarrow} \ldots
\stackrel{\wt\delta}{\longrightarrow} F_1\stackrel{\wt\delta}{\longrightarrow} F_0\,,
\eeq
with
\beq
F_i=\bigoplus_{j=0}^{[{i\over 2}]}K_{i-2j}\otimes T_j\,,\qquad
\wt\delta=\delta\otimes\id+\sigma\otimes \lambda\,.
\eeq
One has $(\wt\delta)^2=W\otimes t$, and since $F_0=K_0$, $F_1=K_1$, we
have $F_0/\im(\wt\delta)=N$. Therefore, tensoring the
complex \eq{Fcomplex} with $R$, we obtain an $R$-free resolution
$\ldots\rightarrow \wt
F_i\rightarrow \wt F_{i-1}\rightarrow\ldots\rightarrow \wt F_0\rightarrow
N\rightarrow 0$ of $N$ with $\wt F_i=F_i\otimes R$. By construction this
complex is $2$-periodic from position $i=n$.

Since the $T$-factors in the periodic part are redundant, the latter can
be represented as follows
\beq\label{perpart}
\Phi_j=\bigoplus_{i}\Lambda^{n-2i+j}V\,,\; j\in\{0,1\}\,,\qquad
\widehat\delta=\delta+\sigma:\Phi_j\longrightarrow\Phi_{j+1}\,.
\eeq
Again $(\widehat\delta)^2=W$, and
$\wt\Phi_i=\Phi_i\otimes R$, together with maps induced by $\widehat\delta$, 
is the periodic part of the $R$-free resolution $\wt F_i$ of $N$.

Now we claim that this periodic part is isomorphic to
the tensor product matrix factorisation 
\beq
M_{L_1,\ldots,L_n}^\otimes=\left(
Q_1\overset{q_1}{\underset{q_0}{\rightleftarrows}} Q_0\right)\,.
\eeq
This can easily be shown by induction on $n$: Let 
$A\p=\CC[x_1,\ldots,x_{n-1}]$, $\Phi_i\p$ and $\widehat\delta\p$ be 
defined as above for the situation with $(n-1)$ variables 
$(x_1,\ldots,x_{n-1})$ and $A\p\p=\CC[x_n]$, $\Phi_i\p\p$ and 
$\widehat\delta\p\p$ be defined as above for the situation with 
one variable $x_n$. Then $\Phi$ is given by the tensor product of 
$\Phi\p$ and $\Phi\p\p$:
$\Phi_i\cong \bigoplus_{r+s+i\in 2\ZZ} \Phi\p_r\otimes_A \Phi\p\p_s$ and 
$\widehat\delta=\widehat\delta\p\otimes\id+\id\otimes\widehat\delta\p\p$. 
Thus, if $(\Phi\p_i,\widehat\delta\p)$ and $(\Phi\p\p_i,\widehat\delta\p\p)$ 
are isomorphic to the respective matrix factorisations, so is 
$(\Phi_i,\widehat\delta)$. Therefore we only have to show the 
statement for the case of one variable, where it is obvious:
\beqn
\Phi_0\cong \Lambda^1 A e_1 \cong \CC[x]\,,&&
\Phi_1\cong \Lambda^0 A e_1 \cong \CC[x]\,,\\
\delta=x^{L+1}:\Phi_0\rightarrow\Phi_1\,,&&
\sigma=x^{d-L-1}:\Phi_1\rightarrow\Phi_0\,.\nonumber
\eeqn
In particular, for a single variable we have $Q_i\cong\wt\Phi_i$.
This proves that the periodic part $\wt\Phi_i$ of the $R$-free 
resolution $\wt F_i$ of $N$ is given by the respective tensor 
product matrix factorisation $M^\otimes_{L_1,\ldots,L_n}$.
However, we need to treat the modules as graded ones, and there 
is a shift of the $\ZZ_d$-grading between $Q_i$ and
$\wt\Phi_i$: Let us assume that the gradings $\alpha$ of the matrix
factorisations are zero. Then, in the one variable case $Q_0$
has degree $0$, whereas $\wt\Phi_0$ has degree $L_1+1$ (since 
this is the degree of the basis vector $e_1$, due to $\delta$ in 
\eq{differentialdeltadef} having degree 0). 
Thus, taking the degrees into account, we find that the tensor product 
matrix factorisation
is isomorphic to the periodic part of the resolution of $N(-\sum_i(L_i+1))$, 
$N$ with degree shifted by $-\sum_i(L_i+1).$\footnote{
If $M = \bigoplus_n M_n$  is a graded module and $\mu$ an integer, 
$M(\mu)$ is the module  with $M(\mu)_n := M_{n-\mu}$. (This is 
not be confused with shifted complexes, usually denoted $C[\mu]$.)
Shifting the degree of a module also affects the degree of its 
Ext-groups, namely $\Ext(M(\mu),N) = \Ext(M,N)(-\mu)$.} 
Let us for the moment abbreviate $\sum_i(L_i+1) =: \mu$. Then we have 
\beq\label{extidsigma}
\Ext_R^i(\coker q_1,M)\cong\Ext^{i+n}_R(N,M)(\mu)
\eeq
for all $i>0$ and all $R$-modules $M$.

To calculate the right hand side of \eq{extidsigma} for 
$M=\coker p_1$, we use the following fact (see \eg 
Lemma 3.1.16 in \cite{Bruns:1993}): Let $S$ be a graded ring, 
$U$ and $V$ be $S$-modules, and $x\in S$ a homogeneous element 
that annihilates $U$ and is $S$- and $V$-regular\footnote{An element $x\in S$ is $V$-regular
if $xv=0$ for $v\in V$ implies $v=0$.}. Then one has 
\beq
\Ext_S^{i+1}(U,V)\cong\Ext_{S/(x)}^i(U,V/xV)(-\deg(x))\,.
\eeq
Noting that $(x_2^{L_2+1},\ldots,x_n^{L_n+1})$ is an $R$- and $\coker p_1$-regular
sequence\footnote{For an $R$-module $M$, an $M$-regular sequence is a sequence
$(a_1,\ldots,a_n)$ in $R$ such that $a_1$ is $M$-regular and $a_{j+1}$ is 
$(M/(a_1,\ldots,a_{j})M)$-regular for all $1\leq j\leq n$.} 
in the annihilator of N, we obtain
\beqn\label{extred}
&&\Ext_R^i(\coker q_1,\coker p_1)\cong\Ext^{i+n}_R\big(N,\coker p_1\big)
({\textstyle \sum_i(L_i+1)})\phantom{***}\\
&&\quad\cong\Ext^{i+1}_{{\scriptstyle R/(x_2^{L_2+1},\ldots,x_n^{L_n+1})}}
\big(N,\coker p_1/{\scriptstyle (x_2^{L_2+1},\ldots,x_n^{L_n+1})}\coker p_1\big) ({\textstyle L_1+1})\,.
\nonumber
\eeqn
The right hand side is easy to determine in the case
$L_2=\ldots=L_n=0$, in which
\beqn 
&&R/(x_2,\ldots,x_n)\ \  \cong\ \  \CC[x_1]/(x_1^d)=:S\,,\nonumber\\
&&\coker p_1/(x_2,\ldots,x_n)\coker p_1\ \ \cong\ \ \left(S/x_1^{d-L-1}S\right)^{d^\gamma}\,,\nonumber\\
&&N\ \ \cong\ \  S/x_1^{L_1+1}S\,,
\eeqn
and $N$ has the obvious $S$-free resolution
\beq
\ldots\longrightarrow S\stackrel{x_1^{L_1+1}}{\longrightarrow} S\stackrel{x_1^{d-L_1-1}}{\longrightarrow} S
\stackrel{x_1^{L_1+1}}{\longrightarrow}S\longrightarrow\CC[x_1]/(x_1^{L_1+1})\longrightarrow 0\,,
\eeq
which can be used to obtain the  
respective $\Ext$-groups  
\beqn
\Ext^{2i+1}_S(N,S/x_1^{d-L-1}S)&\cong&
x_1^{\max(0,L_1-L)}\CC[x_1]/(x_1^{\min(d-L-1,L_1+1)})
({\scriptstyle -L_1-1}) \,,
\nonumber\\
\Ext^{2i}_S(N,S/x_1^{d-L-1}S)&\cong& x_1^{\max(0,d-L-L_1-2)}\CC[x_1]/(x_1^{\min(d-L-1,d-L_1-1)}) \,,
\nonumber
\eeqn
for $i>0$. Putting everything together, we obtain
\beqn
&&\Ext^{2i}_R(\coker q_1,\coker p_1)\cong d^\gamma\left(
x_1^{\max(0,L_1-L)}\CC[x_1]/(x_1^{\min(d-L-1,L_1+1)})\right)\,,\nonumber\\
&&\Ext^{2i+1}_R(\coker q_1,\coker p_1)\nonumber\\
&&\qquad\qquad\qquad\cong d^\gamma\left(
x_1^{\max(0,d-L-L_1-2)}\CC[x_1]/(x_1^{\min(d-L-1,d-L_1-1)})({\scriptstyle
  L_1+1})\right)\ .\nonumber
\eeqn
This agrees, via the correspondence (\ref{correspondencetp},\ref{correspondence}),  precisely
with the topological spectra (\ref{sigmaidtopb},\ref{sigmaidtopf}), and in
particular yields the correct Witten index \eq{eq:wpiid}.

For arbitrary $L_2,\ldots,L_n$, the computation of the right hand side 
of \eq{extred} is more involved. 
Case-by-case checks using Macaulay2 however show agreement in these
cases as well.

\subsection{Computer checks}\label{comtests}

As mentioned above, we have not yet been able to construct a rigorous proof 
for the general correspondence \eq{correspondence}.
Therefore, we collect additional evidence for
it based on case-by-case calculations of
the respective $\Ext$-groups, using the computer algebra program Macaulay2 \cite{M2}.

Macaulay2 does exact calculations using rings which may be of the form
$K[x_1,\ldots,x_n]/{\mathcal I}$, where ${\mathcal I}$ is an ideal and $K$
some field, which we define to be
the field extension $\QQ(a)$, where $a$ is a fundamental
root of $1$ if $d$ is odd and of $-1$ if $d$ is even.
This is done by setting $K=\QQ[a]/(f(a))$ for the appropriate polynomial
$f$.

Macaulay2 has a built-in procedure to calculate the Ext-groups.  The
$\ZZ_d$-representations, which correspond to the degrees of the graded
modules $P_i,Q_i$ and the maps between them, are also calculated by
Macaulay2.  

Although we may use Macaulay2 to calculate the full algebra of the
chiral rings, for brevity we only present the calculation of the 
index $I(P,Q) = \dim\Ext^2(P,Q) - \dim\Ext^1(P,Q)$ here.
Our code can be found in Appendix \ref{code}.
It calculates $I(P,Q)$ for two given graded matrix factorisations $P$
and $Q$ and expresses it in terms of the shift
matrix $G_{\mu\p\mu}$, where here $\mu\p=\alpha\p$ and
$\mu=\alpha$ specify the $\ZZ_d$-representations of $P$ and $Q$
respectively.
A few results are displayed in Appendix \ref{results}.

All our tests showed agreement of the topological spectra of 
permutation boundary conditions on the one hand and the 
graded $\Ext$-groups of the matrix factorisations corresponding to
them via \eq{correspondencetp}, \eq{correspondence}
on the other. 
This is in particular the case for the examples listed in Tables
\ref{inttablepp} and \ref{inttableip}.

\section*{Acknowledgements}
We would like to thank M. Baumgartl, I.\ Brunner, G.\
D'Appollonio, G.\ Ellingsrud, M.\ Gaberdiel, R.\ Ile. W.\ Lerche, 
C.\,A.\ L\"utken and P.\ West for useful discussions. 
This work has been partly supported
by a EC Marie Curie Training Site Fellowship
under contract no.\ HPMT-CT-2001-00296, by the EC 
network ``EUCLID'' under contract no.\ HPRN-CT-2002-00325, and 
by the PPARC grant PPA/G/O/2002/00475. 
We thank the Erwin-Schr\"odinger Institut, Vienna, for hospitality.

\appendix

\section{Deconstructing tensor product factorisations}\label{tpfactapp}

Here we would like to give a slightly different derivation of
\eq{extred}. As in Section \ref{evidence}, we take 
$M^\otimes_{L_1,\ldots,L_n}= (q_0,\,q_1)$ to be a tensor product 
factorisation and $M_{I,I^c}=(p_0,\,p_1)$ to be any linear matrix
factorisation of degree $|I|=L+1$. 
To calculate the modules $\Ext_R(\coker q_1,\coker p_1)$, we can
make use of the tensor product structure of $M^\otimes_{L_1,\ldots,L_n}$.
Namely, 
\begin{equation}\label{tpform}
  q_0 =
\begin{pmatrix}
  q\p_0 \otimes 1 & -1 \otimes q\p\p_1 \\
  1 \otimes q\p\p_0 & q\p_1 \otimes 1
\end{pmatrix},
\quad
  q_1 =
\begin{pmatrix}
  q\p_1 \otimes 1 & 1 \otimes q\p\p_1 \\
  -1 \otimes q\p\p_0 & q\p_0 \otimes 1
\end{pmatrix},
\end{equation}
where 
\beq
\left(
Q\p_1\overset{q\p_1}{\underset{q\p_0}{\rightleftarrows}} Q\p_0\right)
  =M^\otimes_{L_1,\ldots,L_{n-1}}
\eeq
is the tensor product factorisation of 
$W\p(x_1,\ldots,x_{n-1})=x_1^d+\ldots+x_{n-1}^d$ and
$(q\p\p_0=x_n^{L_n+1},q\p\p_1=x^{d-L_n-1})$
the factorisation of $W\p\p(x_n)=x_n^d$.
The  $Q\p_i\otimes Q\p\p_j$ are free, and the long exact $\Ext$-sequence obtained from
\beq
0\longrightarrow Q\p_1\otimes Q\p\p_0\longrightarrow \coker q_0 \longrightarrow
\coker\left( \id_{Q\p_0}\otimes x_n^{L_n+1},\; q\p_1\otimes \id_{Q\p\p_1}\right)
\longrightarrow 0
\eeq
gives rise to the following isomorphisms
\beqn\label{lees}
\Ext^i_R(\coker q_1,\cdot)&\cong&\Ext^{i+1}_R(\coker q_0,\cdot)\\
&\cong&\Ext^{i+1}_R\big( \coker \, q\p_1/x_n^{L_n+1}\coker\,q\p_1,\cdot\big)\,.\nonumber
\eeqn
Following the degrees in all the steps, one sees that the 
degree of the third $\Ext$ in \eq{lees} is shifted
relative to the one of the $\Ext$ on the left hand side by $L_n+1$. 
As in Section \ref{evidence}, we use the fact (see \eg 
Lemma 3.1.16 in \cite{Bruns:1993}) that for any ring $S$ and any
$S$-modules $U$ and $V$ with a homogeneous $x\in S$ that annihilates $U$ 
and is $R$- and $V$-regular
\beq
\Ext_S^{i+1}(U,V)\cong\Ext_{S/(x)}^i(U,V/xV)(-{\rm deg}(x))\,.
\eeq
Since $x_n^{L_n+1}$ is $\coker p_1$-regular this gives 
\beqn
&&\Ext^i_R(\coker q_1,\coker p_1)\\
&&\quad\cong \Ext^i_{R/(x_n^{L_n+1})}\left(\coker q\p_1\big/\,x_n^{L_n+1}\coker q\p_1\,,\  
                     \coker p_1\big/\,x_n^{L_n+1}\coker p_1\;\right)\,.\nonumber
\eeqn
Furthermore, $\coker q\p_1\big/\,x_n^{L_n+1}\coker q\p_1\cong\coker \hat q\p_1$, 
where $\hat q\p_i$ are the induced maps between the $Q_i/x_n^{L_n+1}Q_i$, 
which again have tensor product form \eq{tpform}. Because
$(x_2^{L_2+1},\ldots,x_n^{L_n+1})$ is an $R$- and $\coker p_1$-regular sequence,
we therefore obtain \eq{extred} inductively: 
\beqn
&&\Ext^i_R(\coker q_1,\coker p_1)\nonumber\\
&&\quad\cong
\Ext^i_{R/(x_2^{L_2+1},\ldots,x_n^{L_n+1})}
\left(\wt N,\ \coker p_1\big/{\scriptstyle(
    x_2^{L_2+1},\ldots,x_n^{L_n+1})}\coker p_1\right)\nonumber\\
&&\quad\cong
\Ext^{i+1}_{R/(x_2^{L_2+1},\ldots,x_n^{L_n+1})}
\left(N,\ \coker p_1\big/{\scriptstyle(
    x_2^{L_2+1},\ldots,x_n^{L_n+1})}\coker p_1\right)({\scriptstyle L_1+1})\nonumber
\eeqn
where $N=R/(x_1^{L_1+1},\ldots,x_n^{L_n+1})$ as in Section \ref{evidence} 
and where  we have used $\wt N := R/(x_1^{d-L_1-1},x_2^{L_2+1},\ldots,x_n^{L_n+1})$. 
This provides an alternative derivation of \eq{extred}.

\section{Calculations with Macaulay2}

\subsection{Code}\label{code}

The procedure \texttt{init} sets up the rings necessary for dealing
with linear matrix factorisations of $W=x_1^d+\ldots+x_n^d$.

\footnotesize\begin{verbatim}
-- Sets up necessary fields, rings.
init = (d,n) -> (
     KK=QQ[G]/(1-G^d);
     toField KK;
     K=QQ[a]/((factors(1+(-a)^d))_0);
     toField K;
     K.isHomogeneous=true;
     A=K[x_1 .. x_n];
     f=sum apply(toList(x_1 .. x_n),y->y^d);
     R=A/f;);
\end{verbatim}\normalsize
The procedure \texttt{linmf} creates linear matrix factorisations
\eq{defmf}, where the first argument is an ordered set of indices
labelling the variables used in the factorisation, and the second one 
is the set $I$ defining it. For this procedure we also need some other
functions. 
\footnotesize\begin{verbatim}
-- Function subracts two sets.
subt = (I1,I2) -> (
     I1=I2|I1;
     I1=unique(I1);
     I1=drop(I1,#I2);
     return(I1));

-- Matrices \epsilon_1, \epsilon_2, \epsilon_3
e3mat = (R,d) -> (return(map(R^d,R^d,
                         (i,j)->(if j==(i+1)%d then 1 else 0))));
e1mat = (R,d,v) -> (return(map(R^d,R^d,
                           (i,j)->(if j==(i+1)%d then v^i else 0))));
e2mat = (R,d,v) -> (return(map(R^d,R^d,
                           (i,j)->(if i==j then v^i else 0))));

-- Inductively constructs \alpha matrices
nplustwo = (R,a,I,n,alpha) -> (
     d:=degree R;
     mu:=(if (even ((n-1)//2)) then a else a^(-1));
     if even d then
       (R_(I_(n+1))*e1mat(R,d,a^2))**id_(source alpha)
           +(mu*R_(I_n)*e3mat(R,d))**id_(source alpha)
           +e2mat(R,d,a^2)**alpha
     else
       (R_(I_(n+1))*e1mat(R,d,a))**id_(source alpha)
           +(R_(I_n)*e3mat(R,d))**id_(source alpha)
           +e2mat(R,d,a)**alpha);

alphan = (R,a,I) -> (
     mu:=(if (even ((#I-1)//2)) then a else a^(-1));
     alpha:= (if (even (#I)) then matrix {{mu*R_(I_1)}} 
                             else matrix {{0_R}});
     m:=(if (even (#I)) then 0 else 1);
     for i from (if (even (#I)) then 1 else 0) to floor((#I)/2)-1 do 
       alpha=nplustwo(R,a,I,2*i+m,alpha);
     alpha);

-- Create linear mf
linmf = (I,J) -> (
     d:=degree R;
     a:=((coefficientRing R)_0)_R;
     I=apply(I,i->i-1);
     J=apply(J,i->i%d);
     A=alphan(R,a,I);
     N=rank source A;
     b := if even d then
            z -> (R_(I_0)**id_(R^(rank source A)) + a_R^(2*z) * A)
          else
            z -> (R_(I_0)**id_(R^(rank source A)) + a_R^z * A);
     g=map(R^N,R^N**(R^{#J-d}),product apply(subt(toList(0..d-1),J),b));         
     f=map(source g, target g,(product apply(J,b)));
     return(f,g));
\end{verbatim}\normalsize
The procedure \texttt{tpmf} creates tensor product matrix
factorisations. The first argument is again the ordered set of
variable indices and the second one the ordered set of the respective
$L$-labels.
\footnotesize\begin{verbatim}
-- Creates the tensor product of two matrix factorisations
tp = (p,q) -> (
     Rp0=target p_1;Rp1=source p_1;
     Rq0=target q_1;Rq1=source q_1;
     return(
     map(p_0**id_(Rq0)|-id_(Rp1)**q_1)||(id_(Rp0)**q_0|p_1**id_(Rq1)),
     map(p_1**id_(Rq0)|id_(Rp0)**q_1)||(-id_(Rp1)**q_0|p_0**id_(Rq1))));

-- Creates the tensor product of one-variable factorisations
tpmf = (I,J) -> (
     d=degree R;
     if #I==1 then tf=linmf(I,toList(0..J_0))
              else tf=tp(tpmf(drop(I,-1),J),
                         linmf((I_(#I-1)..I_(#I-1)),toList(0..J_(#I-1))));
     return(tf));
\end{verbatim}\normalsize
The procedure \texttt{deg} calculates the bosonic and fermionic
partition functions and \texttt{ind} the index $I(P,Q)$ between two
matrix factorisations.
\footnotesize\begin{verbatim}
-- Calculates the degrees of the respective Ext-modules
deg = (n,p,q) -> (
     n=-abs(n)%2+2;
     Mp=coker p_1;
     Mq=coker q_1;
     emod=Ext^n(Mp,Mq);
     e=if (dim emod!=0) then matrix {{}} else super basis emod;
     ed=apply(numgens source e,i->((degree e_i)_0));
     return(sum(ed,i->G^(i))));

-- Calculates the index I
ind = (p,q) -> (
     return(deg(0,p,q)-deg(1,p,q)));
\end{verbatim}\normalsize

\subsection{Results}\label{results}

In the following we demonstrate how to use the above code:
\footnotesize\begin{verbatim}
Macaulay 2, version 0.9.2
--Copyright 1993-2001, D. R. Grayson and M. E. Stillman
--Singular-Factory 1.3c, copyright 1993-2001, G.-M. Greuel, et al.
--Singular-Libfac 0.3.3, copyright 1996-2001, M. Messollen

i1 : load "linmf.m2"
--loaded linmf.m2
\end{verbatim}\normalsize
As an example, for $d=4$ and $n=3$, 
we set up a linear matrix factorisation $M_{\{0\}}$ 
and a tensor product factorisation $M^\otimes_{0,1,2}$ and calculate
indices $I$.
\footnotesize\begin{verbatim}
i2 : init(4,3)

i3 : p=linmf({1,2,3},{0});

i4 : p_0

o4 = {3} | x_1        ax_2+x_3 0          0        |
     {3} | 0          x_1      ax_2+a2x_3 0        |
     {3} | 0          0        x_1        ax_2-x_3 |
     {3} | ax_2-a2x_3 0        0          x_1      |

             4       4
o4 : Matrix R  <--- R

i5 : q=tpmf({1,2,3},{0,1,1});

i6 : q_0

o6 = {3} | x_1   -x_2^2 -x_3^2 0      |
     {2} | x_2^2 x_1^3  0      -x_3^2 |
     {2} | x_3^2 0      x_1^3  x_2^2  |
     {7} | 0     x_3^2  -x_2^2 x_1    |

             4       4
o6 : Matrix R  <--- R

i7 : ind(p,p)

         3     2
o7 = - 6G  + 6G  + 2G - 2

o7 : KK

i8 : ind(q,p)

         3     2
o8 = - 4G  + 4G  + 4G - 4

o8 : KK
\end{verbatim}\normalsize
The indices calculated for these matrix factorisations agree via the
correspondence \eq{correspondencetp}, \eq{correspondence} with the
respective entries in Tables \ref{inttablepp}, \ref{inttableip}.

For even cycle length, \eg $n=4$, the factorisations $M_I$ not only depend on
the cardinality of $I$.
\footnotesize\begin{verbatim}
i9 : init(4,4)

i10 : p=linmf({1,2,3,4},{0});

i11 : q=linmf({1,2,3,4},{1});

i12 : ind(p,p)

        3      2
o12 = 4G  + 10G  + 4G + 2

o12 : KK

i13 : ind(p,q)

          3     2
o13 = - 2G  + 4G  - 2G - 4

o13 : KK
\end{verbatim}\normalsize
Upon comparison with Tables \ref{inttablepp} and \ref{inttableip} also
these results agree with the correspondence \eq{correspondence}.

\section{Linear matrix factorisations and the quintic}

\begin{table}
\begin{center}
\small
\begin{tabular}{|c|>{$}c<{$}|>{$}c<{$}>{$}c<{$}>{$}c<{$}>{$}c<{$}|>{$}c<{$}>{$}c<{$}>{$}c<{$}>{$}c<{$}|}
\hline
Permutation & M & \multicolumn{4}{c|}{Charges} &
   \multicolumn{4}{c|}{Chern characters} \\
& & D6 & D4 & D2 & D0 & \rk & \ch_1 & \ch_2 & \ch_3 \\
\hline
(1)(2)(3)(4)(5) & 0 & 1&0&0&0&1&0&0&0\\
& 2 & -4&-1&-8&5&-4&1&5/2&5/6\\
& 4 & 6&3&19&-10&6&-3&-5/2&5/2\\
& 6 & -4&-3&-14&10&-4&3&-5/2&-5/2\\
& 8 & 1&1&3&-5&1&-1&5/2&-5/6\\
\hline
(1)(2)(3)(45) & 0 & 3&2&11&-6&3&-2&0&7/3\\
& 2 & -1&-1&-3&4&-1&1&-5/2&-1/6\\
& 4 & 0&0&0&-1&0&0&0&-1\\
& 6 & 1&0&0&-1&1&0&0&-1\\
& 8 & -3&-1&-8&4&-3&1&5/2&-1/6\\
\hline
(1)(2)(345) & 0 & 0&0&-5&0&0&0&5&0\\
& 2 & -5&0&-5&5&-5&0&5&5\\
& 4 & 10&5&35&-15&10&-5&-15/2&35/6\\
& 6 & -5&-5&-20&15&-5&5&-15/2&-35/6\\
& 8 & 0&0&-5&-5&0&0&5&-5\\
\hline
(1)(23)(45) & 0 & 0&0&-1&0&0&0&1&0\\
& 2 & -1&0&-1&1&-1&0&1&1\\
& 4 & 2&1&7&-3&2&-1&-3/2&7/6\\
& 6 & -1&-1&-4&3&-1&1&-3/2&-7/6\\
& 8 & 0&0&-1&-1&0&0&1&-1\\
\hline
(1)(2345) & 0 & 5&4&22&-10&5&-4&0&20/3\\
& 2 & 0&-1&2&5&0&1&-15/2&5/6\\
& 4 & 0&-1&-3&0&0&1&-5/2&-25/6\\
& 6 & 0&-1&-8&0&0&1&5/2&-25/6\\
& 8 & -5&-1&-13&5&-5&1&15/2&5/6\\
\hline
(12)(345) & 0 & 5&4&22&-10&5&-4&0&20/3\\
& 2 & 0&-1&2&5&0&1&-15/2&5/6\\
& 4 & 0&-1&-3&0&0&1&-5/2&-25/6\\
& 6 & 0&-1&-8&0&0&1&5/2&-25/6\\
& 8 & -5&-1&-13&5&-5&1&15/2&5/6\\
\hline
(12345) & 0 & -5&0&-25&0&-5&0&25&0\\
& 2 & -5&5&15&0&-5&-5&25/2&125/6\\
& 4 & 20&10&80&-25&20&-10&-25&50/3\\
& 6 & -5&-10&-30&25&-5&10&-25&-50/3\\
& 8 & -5&-5&-40&0&-5&5&25/2&-125/6\\
\hline
\end{tabular}
\caption{Charges and Chern characters of $L=0$ permutation branes for
the quintic, computed from the Witten index with tensor product branes}
\label{table:chern}
\end{center}
\end{table}
\normalsize

For the quintic hypersurface in $\PP^4$, the charges of the minimal
model tensor product branes were calculated in
\cite{Brunner:1999jq}. The rank of the B-brane charge lattice is equal
to $N=\sum_i b^{2i}$, where $b^j$ are the Betti numbers of the
underlying manifold; $N=4$ for the quintic.
 
Since the charges of the $L_i=0$ tensor product branes span the whole
charge lattice over $\QQ$ in this case (they do not provide an
integral basis), one can extract the charges of the permutation branes
from the Witten index $I^{\id\,\sigma}$ between $L_i=0$ tensor product
branes and the permutation branes determined in \eq{iidsigma}: Let the
columns of the matrix $Q^\id$ contain the charges of the $L_i=0$
tensor product branes\footnote{The charge matrix $Q^\id$ for a general
Gepner model may be determined from the results of
\cite{Mayr:2000as}}, and the ``large volume intersection matrix'' $I$
in the charge basis be given by $I_{ij}=(-1)^{i+1}\delta_{i,n-i+1}$.
As long as $Q^\id$ has rank $N$ (meaning that the tensor product
branes span the whole charge lattice over $\QQ$), the charges of
$\sigma$-permutation branes are given by $Q^\sigma=J I^{\id\,\sigma}$,
where $J (Q^\id)^t I = 1$.

By the method outlined above, we can compute all permutation brane 
charges from the intersection form $I^{\id\,\sigma}$ alone, for any 
model where the tensor product branes generate the charge lattice over $\QQ$. 
The resulting charges for all B-type permutation branes with
$L$-labels $0$ on the quintic 
are displayed in Table \ref{table:chern}. Note that if the
  charge lattice is generated (over $\QQ$) by the tensor product branes,
  it easily follows from the form of $I^{\id\,\sigma}$ given in
  \eq{iidsigma} that the charges of the permutation branes with $L=0$
  generate those of all permutation branes.

Among the branes for the permutation $(12)(3)(4)(5)$, one finds one 
with charges of a (single) D0-brane. The absence of other charges 
(D2,4,6) for this boundary state was already noted in \cite{r02,r03}, 
where however the normalisation was not discussed. The correct
normalisation was first obtained in \cite{Brunner:2005fv}, where it
was noticed that the charges of the $(12)(3)(4)(5)$ permutation branes
indeed generate the whole charge lattice (over the integers). 

Another interesting property to note is that, 
up to normalisation, the intersection forms \eq{iidsigma} only depend on the number of 
cycles of the respective permutation. The normalisation is given by $(k+2)^N$
with $N=\sum_\nu [{n_\nu-1\over 2}]$ depending on the cycle lengths
$n_\nu$ only.
In particular the normalisation for the permutation branes $(12)(345)$ and
$(1)(2345)$ and therefore their charges are identical. Only for these
permutation branes D4 branes (without D6-brane charge) show up.  
None of these boundary states, however, is a ``pure'' D4 brane, instead 
there is some admixture of D2-charge.

\providecommand{\href}[2]{#2}\begingroup\raggedright\endgroup

\end{document}